\documentclass[8.5pt,twoside]{article}

\usepackage[super,sort&compress,comma]{natbib} 
\usepackage{times,mathptm}
\usepackage{sectsty}
\usepackage{balance} 
\usepackage[text={11.3cm,20.4cm},centering]{geometry} 
\usepackage{color}
\usepackage{amsfonts}
\usepackage{amsmath}
\usepackage{amssymb}

\usepackage{graphicx} 
\usepackage{lastpage}
\usepackage[format=plain,justification=raggedright,singlelinecheck=false,font=small,labelfont=bf,labelsep=space]{caption} 
\usepackage{fancyhdr}
\newtheorem{theorem}{Theorem1}[section]

\newcommand{\qed}{\nobreak \ifvmode \relax \else
      \ifdim\lastskip<1.5em \hskip-\lastskip
      \hskip1.5em plus0em minus0.5em \fi \nobreak
      \vrule height0.75em width0.5em depth0.25em\fi}

\begin{document}

\thispagestyle{plain}
\fancypagestyle{plain}{
\fancyhead[L]{\includegraphics[height=8pt]{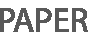}}
\fancyhead[C]{\hspace{-1cm}\includegraphics[height=15pt]{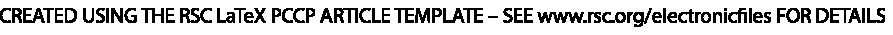}}
\fancyhead[R]{\includegraphics[height=10pt]{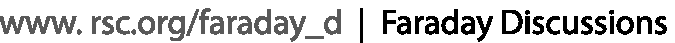}\vspace{-0.2cm}}
\renewcommand{\headrulewidth}{1pt}}
\renewcommand{\thefootnote}{\fnsymbol{footnote}}
\renewcommand\footnoterule{\vspace*{1pt}%
\hrule width 11.3cm height 0.4pt \vspace*{5pt}} 
\setcounter{secnumdepth}{5}

\makeatletter 
\renewcommand{\fnum@figure}{\textbf{Fig.~\thefigure~~}}
\def\subsubsection{\@startsection{subsubsection}{3}{10pt}{-1.25ex plus -1ex minus -.1ex}{0ex plus 0ex}{\normalsize\bf}} 
\def\paragraph{\@startsection{paragraph}{4}{10pt}{-1.25ex plus -1ex minus -.1ex}{0ex plus 0ex}{\normalsize\textit}} 
\renewcommand\@biblabel[1]{#1}            
\renewcommand\@makefntext[1]%
{\noindent\makebox[0pt][r]{\@thefnmark\,}#1}
\makeatother 
\sectionfont{\large}
\subsectionfont{\normalsize} 

\fancyfoot{}
\fancyfoot[LO,RE]{\vspace{-7pt}\includegraphics[height=8pt]{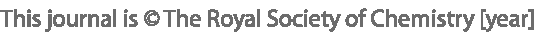}}
\fancyfoot[CO]{\vspace{-7pt}\hspace{5.9cm}\includegraphics[height=7pt]{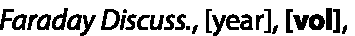}}
\fancyfoot[CE]{\vspace{-6.6pt}\hspace{-7.2cm}\includegraphics[height=7pt]{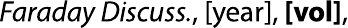}}
\fancyfoot[RO]{\scriptsize{\sffamily{1--\pageref{LastPage} ~\textbar  \hspace{2pt}\thepage}}}
\fancyfoot[LE]{\scriptsize{\sffamily{\thepage~\textbar\hspace{3.3cm} 1--\pageref{LastPage}}}}
\fancyhead{}
\renewcommand{\headrulewidth}{1pt} 
\renewcommand{\footrulewidth}{1pt}
\setlength{\arrayrulewidth}{1pt}
\setlength{\columnsep}{6.5mm}
\setlength\bibsep{1pt}


\noindent\LARGE{\textbf{Nonlinear quantum optics
\\
in the (ultra)strong light-matter coupling}}
\vspace{0.6cm}


\noindent
\large{\textbf{Eduardo S\'anchez-Burillo\textit{$^{a}$}, 
Juanjo Garc\'ia-Ripoll\textit{$^{b}$}, \\
Luis Mart\'in-Moreno \textit{$^{a}$}
 and David Zueco\textit{$^{a, c}$}}}\vspace{0.5cm}

\vspace{0.6cm}


\noindent \normalsize
{
The propagation of $N$ photons in one dimensional waveguides coupled to $M$
qubits is discussed, both in the strong and ultrastrong qubit-waveguide coupling.  Special emphasis is placed on the characterisation of the nonlinear response and
its linear limit  for
the scattered photons as a function of $N$, $M$, qubit inter distance
and light-matter coupling.  
The quantum evolution is numerically solved
via the Matrix Product States technique.
 Both the
time evolution for the field and qubits is computed.
The nonlinear character (as a function of $N/M$) depends on the
computed observable.  While perfect reflection is obtained for $N/M
\cong 1$,  photon-photon correlations are still
resolved for ratios $N/M= 2/20$.
Inter-qubit distance enhances the nonlinear response.
Moving to the ultrastrong coupling regime,  we observe that inelastic
processes are \emph{robust} against the number of qubits
and that the qubit-qubit interaction mediated by the photons is 
qualitatively modified. 
The theory  developed in this work modelises experiments in circuit QED,
photonic crystals and dielectric waveguides.
}
\vspace{0.5cm}



\section{Introduction}
\label{sec:intro}


\footnotetext{\textit{$^{a}$
Instituto de Ciencia de Materiales de Arag\'on y Departamento de F\'{\i}sica de la Materia Condensada, CSIC-Universidad de Zaragoza, Zaragoza, E-50012, Spain.}}

\footnotetext{\textit{$^{b}$Instituto de F\'{\i}sica Fundamental,
    IFF-CSIC, Serrano 113-bis, Madrid E-28006, Spain }}

\footnotetext{\textit{$^{b}$
Fundaci\'on ARAID, Paseo Mar\'{\i}a Agust\'{\i}n 36, Zaragoza 50004, Spain
}}

Typically, materials respond linearly to the electromagnetic (EM)
field.
Intense fields are usually demanded for accessing the nonlinear
response \cite{Boyd2003}.  Therefore, a long standing challenge in science and
technology is to develop devices containing giant nonlinear properties
at small powers.  The final goal is to shrink the required power to the few
photon limit \cite{Lukin2012b, Lukin2013}.  
In doing so, the dipoles must interact
more strongly with the driving photons than with the environment, which defines 
the \emph {strong light-matter coupling} regime.
Thus, quantum optical systems presenting
strong light matter interaction are excellent candidates  for
building  nonlinear optical materials operating at tiny powers.

An ideal platform for having strong light-dipole coupling together with
the possibility of generating and measuring few photon currents is   waveguide QED. 
There, the paradigmatic dipoles are two level systems (qubits) and
the input and output fields travel through one dimensional waveguides. 
As there are only two propagation directions (left and right), interference effects are much larger than in 3D.
Besides, the  
coupling to the qubits is enhanced by the reduced dimensionality
(Purcell effect).
Different platforms can serve for the study: circuit-QED
\cite{You2011, Astafiev2010,Johansson2013b,Wallraff2013b}, quantum
dots interacting with photonic crystals\cite{Lodahl2008}, dielectric
waveguides\cite{OBrien2007}, molecules interacting with
photons\cite{Sandoghdar2012} or plasmonic
devices\cite{Maier2007,Lukin2006,Lukin2007b,Lukin2009, Lukin2012a}.
With this kind of systems different nonlinear effects  may be observed and used, as
photon-photon correlations \cite{Fan2007a,Fan2007b,Fan2011a,Fan2011b,Fan2013,Sun2009a,Reineker2008,Baranger2010,Baranger2011,
Baranger2012,Baranger2013,Baranger2014,Roy2011b,Roy2013,Liao2010}, non-classical light generation \cite{Johansson2013a}, lasing \cite{Fan2012} or effective
interaction between noninteracting dipoles
\cite{Nori2011,Tudela2011a,Zueco2012, Wallraff2013a,Ballestero2014}.

Light-matter coupling, even when it is larger than the losses, 
is typically  much smaller than the
characteristic energy scales of qubits (dipoles) and photons.
In this case, up to second order in this coupling, only  processes where light and
matter exchange excitations play a role.  This is the Rotating Wave
Approximation (RWA) \cite{Cohen-Tannoudji1992}.
Quite recently, experiments have reached couplings large enough for this
approximation to break down \cite{Niemczyk2010,FornDiaz2010,Schwartz2011,Gunter2009}. 
Then, the full dipole interaction must
be taken into account and, in order to understand the experiments, it is indispensable to consider processes involving spontaneous creation
and annihilation of pairs of excitations.
This is the 
ultrastrong coupling regime.
Beyond the RWA picture novel physics has been predicted
\cite{DeLiberato2008,Lehur2012,Romero2012,Peropadre2013,Naether2014},
also from the scattering point of view \cite{Burillo2014, Sun2012}. 
Clearly, this regime has a great potential for nonlinear applications.

It is desirable to quantify the
amount of \emph{nonlinearity} for a given architecture with a
given input driving, like in \emph{classical} nonlinear
optics, where materials are classified via their linear and nonlinear susceptibilities.  
Furthermore, some systems can  behave  as
linear when looking at one quantity and nonlinear when measuring
another.
In general, the response is expected to be
linear in the low polarisation limit, $N/M \ll 1$, with $N$ the number of
photons and $M$ the number of qubits.
 In this work, we quantify more precisely this linear-nonlinear transition. 
In doing so, we must notice that the description of qubits and 
currents containing few photons needs a quantum treatment.
To compute this fully quantum evolution, we 
choose the Matrix Product State (MPS) technique adapted for
photonic situations \cite{Peropadre2013, Burillo2014}.  
Within the MPS, the \emph{exact} dynamics is computed for  multiphoton input
states passing through several qubits both in the strong and
ultrastrong coupling regimes.
We explore the dynamics by  changing the ratio $N/M$ and the
light-matter coupling strength.
Our aim here is to discuss the tradeoff between the enhancement of the
coupling with the number of qubits (which, accordingly to the theory of Dicke states\cite{Dicke1954}
scales as $g \sqrt{M}$) at the expense of decreasing the nonlinear response.
Besides, we also study the influence of the distance between the dipoles.
%
Finally, we will explore the changes on this RWA phenomenology when
moving to the  ultrastrong coupling regime.
As witnesses for the nonlinearities we will compute transmission and
reflection probabilities, qubit populations and photon-photon correlations.

The rest of this \emph{Discussions} is organised as follows. 
Section \ref{sec:TB} summarises the theoretical basis needed for
understanding the results.  Section \ref{sec:linear} discusses the
analytic properties for the scattering matrix in linear systems (i.e. when the
scatterers are harmonic resonators), linking the notion of linear
quantum system with linear optical response.  The numerical results both
in the RWA and beyond the RWA are presented in Sects. \ref{sec:results}
and \ref{sec:ultrastrong} respectively.  Our conclusions and two
technical Appendices close the paper.

\section{Theoretical background}
\label{sec:TB}

\begin{figure}[t]
\includegraphics[width=1.\columnwidth]{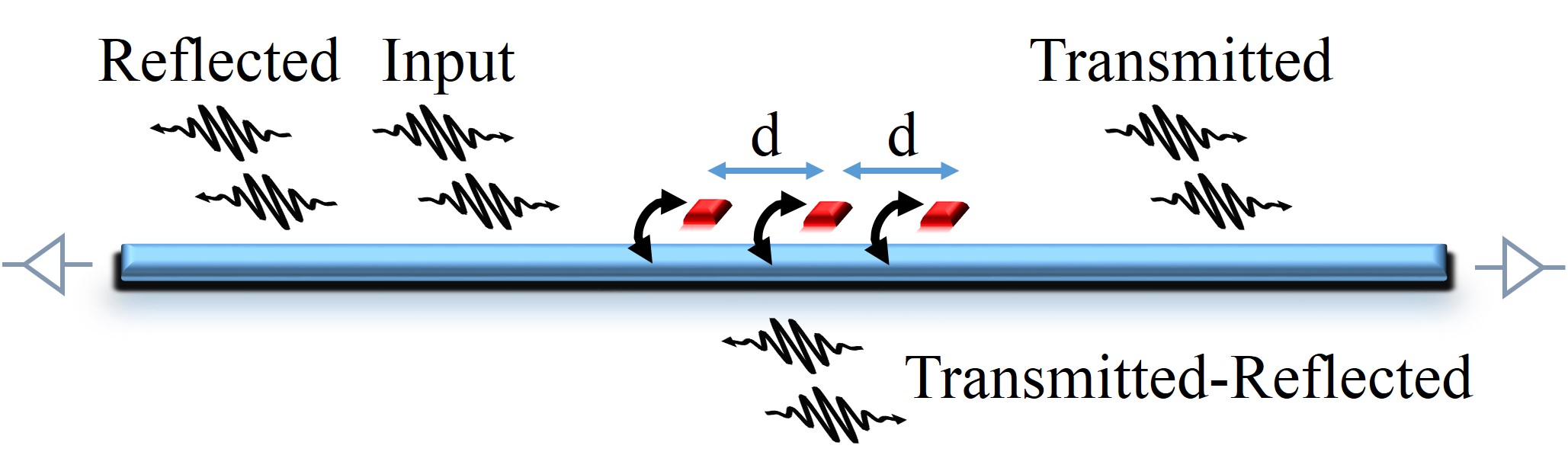}
\includegraphics[width=1.\columnwidth]{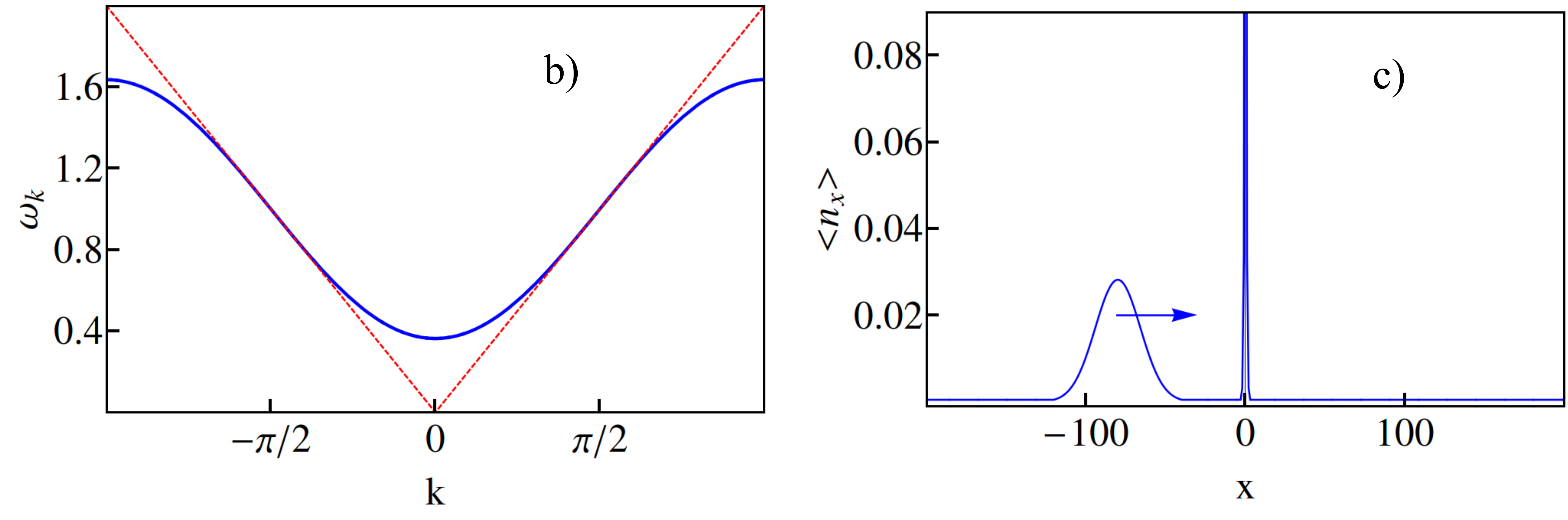}\\
\caption{{\it (Colour online)} a) Schematic representation of the
  system: waveguide with input and output fields and the qubits (red
  squares) separated by a distance $d$. b) Dispersion relation (blue curve) for
  $\epsilon= 1$ and $J=1/\pi$ and $400$ cavities (these parameters will be used throughout
  the text) .  The dashed red line is the linearisation of $\omega_k$
  around $k=\pi/2$. c) Computed number of photons as a function on the
  chain 
  (\ref{eq:nx}) for an incident wavepacket ($N=1$) impinging on one qubit
  ($\Delta=1$) taking the full interaction Hamiltonian
  (\ref{eq:Htot}) with $g=0.7$. There is a photonic cloud around the qubit position ($x=0$), corresponding to the ground state. This ground state was computed with the MPS method, as explained in section \ref{sec:mps_method}.}\label{fig:fig_intro}
\end{figure}

We summarise the quantum theory for scattering in
waveguide QED.    We also sketch the MPS technique used in our
numerical simulations.

\subsection{Physical setup}
\label{sec:physical}

This work deals with the dynamics of photons moving in a one
dimensional waveguide interacting with a discrete number of dipoles. 
A pictorial representation is given in Fig. \ref{fig:fig_intro}~a).  
The waveguide is discretised and the quantum dipoles (scatterers) are point like
  systems interacting with the photons accordingly to the dipole approximation,
$H_{int}\propto \mathbf{E}\cdot\mathbf{p}$. 
The Hamiltonian is,
\begin{equation}
\label{eq:Htot}
H_{\rm tot}=\epsilon\sum_x a_x^\dagger a_x - J \sum_x (a_x^\dagger
a_{x+1} + {\rm  hc} ) + \sum_{i=1}^M \left(h_i(c_i,c_i^\dagger) + g_i (c_i +
  c_i^\dagger)(a_{x_i}+a_{x_i}^\dagger)\right) \, .
\end{equation}
Here, the waveguide is modeled within the first two terms.  It is
an array of $N_{cav}=2L+1$ coupled cavities, running $x$ from $-L$ to $L$, with bare frequencies $\epsilon$ and hopping between nearest neighbours
$J$. Photon operators satisfy bosonic relations $[a_x, a_{x'}^\dagger]= \delta_{xx'}$. The free propagation is characterised by a cosine-shaped dispersion relation $\omega_k = \epsilon - 2 J \cos k$, as shown
in figure \ref{fig:fig_intro}~b).   
Under some
circumstances and working
in the middle of the band $k \cong \pi/2$,  the linearised dispersion
$\omega_k = v k$ with $v= 2 J$ is well justified.
Such a linearisation
is not performed in our numerical investigations, but it can be
important to have it in mind when comparing some of our findings with analytical treatments (mostly
done with
linear dispersion relations).
The dipoles are characterised via the ladder operators $c_i^\dagger$
($c_i$). 
In the numerical work presented in this paper, we will consider qubits,
\begin{equation}
\label{c-qubit}
c_i^\dagger=\sigma_i^+
\; , {\rm with} \; ,
[\sigma_i^+,
\sigma_j^-] = \delta_{ij} ( 2 \sigma^+_i \sigma^-_i - 1)
\end{equation}
and
\begin{equation}
\label{h-qubit}
h_i(c_i,c_i^\dagger)=\Delta_ic_i^\dagger c_i
\end{equation}
 with  $\Delta_i$  the frequency for each qubit.
Other nonlinear dipoles could be investigated without extra
difficulty \emph{e.g.}  three level atoms \cite{Sorensen2010,Roy2011a} or Kerr-type punctual materials\cite{Liao2010}. We can also consider linear dipoles, i.e., harmonic oscillators satisfying bosonic commutation relations $[c_i,c_j^\dagger]=\delta_{ij}$, as we will do in section \ref{sec:lin-nolin} in order to explore the linear limit.

The last term in (\ref{eq:Htot})  results from the quantisation of the
dipole interaction term $(H_{int})_i\propto \mathbf{E}(x_i)\cdot\mathbf{p}_i$,
because
$E(x_i) \sim a_{x_i}+ {\rm h.c.}$\cite{Cohen-Tannoudji1997} and $p_i \sim c_i^\dagger +{\rm h.c.}$ \cite{Cohen-Tannoudji1992}.  
The coupling strength is given by the constants $g_i$, whose actual
value will depend on the concrete physical realisation.  
If $g_i \ll  \Delta_i$ we can approximate the interaction as,
\begin{equation}
\label{eq:hintrwa}H_{int}=g\sum_{i=1}^M(c_i+c_i^\dagger)(a_{x_i}+a_{x_i}^\dagger)\simeq g\sum_{i=1}^M(c_i^\dagger a_{x_i}+c_i a_{x_i}^\dagger)\equiv H_{int}^{RWA}\, .
\end{equation}
{\it i.e.} we have neglected the counter-rotating terms
$c_i^\dagger a_{x_i}^\dagger + {\rm h.c.}$. 
This is the so-called Rotating Wave Approximation (RWA), which
is valid up to ${\mathcal O} (g_i^2)$
\cite{Cohen-Tannoudji1992}.  It is widely
used since  $g_i\ll  \Delta_i$ typically holds in
 the experiments.
Within the RWA,   the number of
excitations is conserved, 
 $[H,\mathcal{N}]=0$, being $\mathcal{N}\equiv \sum_x a_{x}^\dagger
 a_{x} + \sum_i c_i^\dagger c_i$ the number operator.
This symmetry highly reduces the  complexity of solving
the dynamics of (\ref{eq:Htot}).

If $g_i$ is not  a small parameter, the RWA is not
justified and the full dipole interaction [last term in
(\ref{eq:Htot})] must be taken into account.  
The regime where the RWA is
not sufficient for describing  the phenomenology is known as the  {\it ultrastrong}
coupling regime
\cite{Gunter2009, Niemczyk2010,FornDiaz2010, Schwartz2011}
Here, the number of excitations is not conserved,
making the problem much harder to solve. 

A good  example to feel the
extra complexity, is the computation of the ground state for (\ref{eq:Htot}),
which is an essential step in the scattering process. Within the RWA, the ground state is trivial,
$|GS\rangle_{RWA} = |0\rangle$, with
$a_{x} |0 \rangle = 0$ for all $x$ and $c_j |0\rangle = 0 $ for all $j$.
Beyond the RWA, in the ultrastrong coupling regime, the $|GS\rangle$
is a correlated state with
$\langle GS | \mathcal {N} | GS \rangle \neq 0$. 
In figure \ref{fig:fig_intro} c) the photon population in the waveguide
per site, $\langle n_x  \rangle=\langle \Psi |a_{x}^\dagger a_{x} | \Psi \rangle$ is
plotted, being $|\Psi\rangle$ the ground state with a flying photon (Eq. \ref{eq:psi_in} for $N=1$). There, only one qubit is coupled to the waveguide beyond
the RWA ($g=0.7 [\Delta]$).  Around the qubit
(placed at $x=0$) a non trivial structure appears. The peaked
wavepacket around $x=-90$ is the flying photon.


\subsection{Scattering}
\label{sect:scattering}

We are interested in computing the scattering characteristics  for
$N$-photon input states  interacting with $M$  qubits.  
The input state,  our initial condition, is chosen to be the non-normalised quantum state,
\begin{equation}
\label{eq:psi_in}|\Psi_{in}\rangle = 
(a_\phi^\dagger)^N|GS\rangle,\qquad a_\phi^\dagger = \sum_x \phi^{in}_x a_x^\dagger\, ,
\end{equation}
where $|GS\rangle$ is the ground state of the system [See
Fig. \ref{fig:fig_intro}c)]
 and $\phi_x^{\rm in}$ is a Gaussian wavepacket centred in $x_{\rm in}$
 with spatial width $\theta$
\begin{equation}
\label{eq:wp}
\phi_x^{in}=\exp\left(-\frac{(x-x_{\rm
      in})^2}{2 \theta ^2}+ik_{in}x\right)
\; .
\end{equation}
Typically,  we consider $x_{\rm in}$ located on the left hand side of the qubits with the wavepacket moving to the right toward the scatterers, as sketched in
Fig. \ref{fig:fig_intro} a and computed in Fig. \ref{fig:fig_intro} c.

In several occasions, it will be convenient to work in
momentum space,
\begin{equation}
\label{eq:m-p}
a_k^\dagger = \frac{1}{\sqrt{L}}
\; 
\sum_x {\rm e}^{i k x} a_x^\dagger
\; .
\end{equation} 
In momentum space the wavepacket (\ref{eq:wp}) is  exponentially
localised  around $k_{in}$, with width $\sim\theta^{-1}$.

In our numerical simulations we evolve the initial state (\ref{eq:psi_in}) unitarily,
\begin{equation}
| \Psi (t) \rangle = U(t, 0) |
\Psi_{\rm in} \rangle = {\rm e}^{-i H_{\rm tot} t} | \Psi_{\rm in} \rangle 
\; .
\end{equation}
We stop the simulation at a final time $t_{out}$, which must be sufficiently
large for the photons to be moving freely along the waveguide,
after having interacted with the scatterers.
In doing so,  our numerics account for
stationary amplitudes encapsulated in the definition of the
scattering matrix, $S$,
\begin{equation}
\label{eq:psi_out}|\Psi_{out}\rangle = S|\Psi_{in}\rangle\, .
\end{equation}
It is customary to specify the scattering matrices through their
momentum components:
\begin{equation}
\label{eq:Spk}
S_{p_1  ...  p_{N^{\prime}} \, , \, k_1 ...  k_N}
=
\langle GS  | a_{p_1} ... a_{p_{N^\prime}}  \, S  \, a_{k_1}^\dagger
... a_{k_{N}}^\dagger  | GS \rangle
\; .
\end{equation}
Some comments are pertinent here.  
The $|GS\rangle$ appears in the definition of $S$.  As discussed before, in the
ultrastrong coupling regime the ground state differs from the vacuum state, and has a non-zero number of excitations, see figure \ref{fig:fig_intro} ~c).  In the ultrastrong regime the number
of excitations is not conserved. Therefore, in the above formula $N^\prime \neq
N$ in general. The components, $S_{p_1  ...  p_{N^{\prime}} \, , \, k_1 ...  k_N}$
can be numerically computed as projections of the evolved state as we
will explain below.


\subsection{Matrix Product States for scattering problems}
\label{sec:mps_method}

As said, beyond RWA even the ground state is non trivial. The problem
becomes a many-body one and we must consider the full Hilbert space
for any computation. If we truncate the number of particles per site
to $n_{max}$ and our scatterers are qubits, the dimension of the
Hilbert space is $2^M (n_{max}+1)^{N_{cav}}$, which is exponential with both
$N_{cav}$ and $M$. Numerical brute force is impossible and analytical
tools are really limited. Even within the RWA, if we work with $N$
photons and $M$ qubits the Hilbert space dimension, $(N_{cav}+M)^N$, is also
too large for multiphoton states.

In order to solve the problem, we use
 one celebrated method to deal with many-body 1D problems: Matrix Product States (MPS)\cite{Vidal2003, Vidal2004, Verstraete2008}. Let us summarise the idea behind this technique. A general state of a many-body system is usually written as
\begin{equation}
\label{eq:psi}|\Psi\rangle = \sum_{i_i,\dots,i_{N_{cav}}} c_{i_1,\dots,i_{N_{cav}}}|i_1,\dots,i_{N_{cav}}\rangle\, ,
\end{equation}
where $\{|i_n\rangle\}_{i_n=1}^{d_n}$ is a basis of the local Hilbert
space of the $n$-th body (or site, in our case), being $d_n$ the
dimension of this local Hilbert space,
$|i_i,\dots,i_{N_{cav}}\rangle=|i_1\rangle\dots|i_{N_{cav}}\rangle$, that is, the
tensor product basis, and $c_{i_1,\dots,i_{N_{cav}}}\in\mathbb{C}$. However,
it is possible to show that an equivalent parametrisation  can be written as:
\begin{equation}
\label{eq:mps}|\Psi\rangle = \sum_{i_i,\dots,i_{N_{cav}}} A_1^{i_1}\dots A_{N_{cav}}^{i_{N_{cav}}} |i_1,\dots,i_{N_{cav}}\rangle\, ,
\end{equation}
where $A_n^{i_n}$ is a $D_n\times D_{n+1}$ matrix linked to the $n$-th
site\footnote{We are taking open boundary conditions, so
  $D_{{N_{cav}}+1}=D_1=1$.}.
For the sake of simplicity, let us take 
$d_n=d$  and $D_n=D$ for all $n$. Then, $N_{cav}D^2d$
coefficients are needed to describe any state. 
In principle, to represent $|\Psi\rangle$ exactly, $D$ must be
exponential in $N_{cav}$. 
However,
states in the low energy sector of well-behaved many-body
Hamiltonians can be very accurately described
by taking $D$ polynomial in $N_{cav}$\cite{Eisert2010}.
Few photon dynamics belongs to this low energy sector and, as we will
show, the number of  MPS coefficients  required to study scattering scales polynomially with $N_{cav}$.
This technique has been recently applied to propagation in bosonic chains interacting with impurities\cite{Peropadre2013,Burillo2014}.

Our simulations are as follows: (i) initialisation of the state as
$|0\rangle$, (ii) computation of the ground state by means of
imaginary time evolution, using the Suzuki-Trotter
decomposition\cite{Suzuki1991} adapted to the MPS representation \cite{JJRipoll2006}, (iii) generation of the input state
(Eq. \ref{eq:psi_in}), (iv) time evolution of the wave function up to
$t=t_{out}$ (Eq. \ref{eq:psi_out}) again by means of Suzuki-Trotter
decomposition and (v) computation of relevant quantities.

\subsection {Observables and its computation with MPS}
\label{sec:observables}

The time dependence of expected values,
\begin{equation}
\langle {\mathcal O}^t \rangle = \langle \Psi(t)| {\mathcal O}
|\Psi(t) \rangle
\label{O}
\end{equation}
with ${\mathcal O} $ any Hermitian operator acting on waveguide, qubit variables or
both, characterises completely the dynamics.
Relevant values are the number of photons (in position and momentum spaces):
\begin{align}
\label{eq:nx} \langle n_x^t\rangle =  \langle \Psi(t)| a_x^\dagger
a_x|\Psi(t) \rangle
\; 
\longleftrightarrow
\; 
 \langle n_k^t\rangle =  \frac{1}{L} \sum_{x_1x_2} e^{ik(x_1-x_2)}\langle\Psi(t)| a_{x_1}^\dagger a_{x_2}|\Psi(t)\rangle,
\end{align}
or the qubit populations and correlations:
\begin{equation}
\label{eq:qubit}P_{i}^t = \langle \Psi(t)|\sigma_i^+\sigma_i^-|\Psi(t)\rangle\, ,\qquad \langle \sigma_i^+ \, \sigma_j^- \rangle
=
\langle \Psi(t)|   \sigma_i^+ \, \sigma_j^- |\Psi(t) \rangle\, .
\end{equation}
Thus, for example, transmission and reflection coefficients can be obtained as
\begin{align}
\label{eq:RT}
T_k =  \langle n_{k}^{t_{out}}\rangle/\langle n_k^0\rangle\, ,
\qquad
R_k =  \langle n_{-k}^{t_{out}}\rangle/\langle n_k^0\rangle\, .
\end{align}
The projectors
\begin{equation}
\label{eq:phiN}\phi_{x_1,\dots,x_N}^t = \frac{1}{\sqrt{N!}}\langle GS|a_{x_1}\dots a_{x_N}|\Psi(t)\rangle\, ,
\end{equation}
are fundamental, since, setting $t=t_{out}$ they are nothing but the
Fourier transform of the scattering matrix $S_{p_1  ...
  p_{N^{\prime}} \, , \, k_1 ...  k_N}$, Eq.  (\ref{eq:Spk}).

Let us start by considering an operator that can be expressed as product of local operators:
\begin{equation}
\mathcal{O}=o_1 \,  o_2 \, \dots \, o_{N_{cav}}
\end{equation}
\emph{e.g.} $ O = a_{x_i}^\dagger \, a_{x_j} $ or $O = \sigma^-_{j}
\, a_{x_j}$.
For the states $|\Psi_A\rangle$ and $|\Psi_B\rangle$,
characterised by the tensors $\{A_n^{i_n}\}_{n=1}^{N_{cav}}$ and
$\{B_n^{i_n}\}_{n=1}^{N_{cav}}$ respectively [Cf. Eq. 
(\ref{eq:mps})] the matrix element
$\langle\Psi_A|\mathcal{O}|\Psi_B\rangle$ is given by 
\begin{equation}
\label{eq:matrix_element}\langle \Psi_A| \mathcal{O}|\Psi_B\rangle = \prod_{n=1}^{N_{cav}} E_n(A_n,B_n,o_n),
\end{equation}
with
\begin{equation}
\label{eq:En}E_n(A_n,B_n,o_n)=\sum_{i_n,j_n} \langle i_n|o_n|j_n\rangle \left( (A_n^{i_n})^* \otimes B_n^{i_n}\right) 
\end{equation}
Since any  operator can always be written as a sum of products of
local operators, we can compute any expected value (\ref{O}) and
projector (\ref{eq:phiN}) without the explicit computation of  the
$c_{i_1,\dots,i_{N_{cav}}}$ coefficients  in Eq. (\ref{eq:psi}).

Let us specify the concrete MPS formulas for the relevant
observables.  If we are interested in the photon number, 
 $\langle n_x^t\rangle$ (\ref{eq:nx}), we compute
 (\ref{eq:matrix_element}) with $|\Psi_A\rangle = |\Psi_B\rangle =
 |\Psi(t)\rangle$, $o_x = a_x^\dagger a_x$, and $o_n=\mathbb{I}_n$
 for $n \neq x_i$, with $\mathbb{I}_n$ the identity operator.
 In the same way, for $P_i^t$ defined in (\ref{eq:qubit}) we must
 replace $o_{x_i} = \sigma_i^+ \sigma_i^-$ and $o_n = \mathbb{I}_n$
 for $n\neq x_i$.
The momentum occupation number, $\langle n_k^t\rangle$ in
(\ref{eq:nx})  is
computationally harder.
We must compute every two-body operators $\langle a_{x_1}^\dagger
a_{x_2}\rangle$ by taking $o_{x_1}=a_{x_1}^\dagger$, $o_{x_2} =
a_{x_2}$ and $o_n = \mathbb{I}_n$ for $n\neq x_1,x_2$, with
$|\Psi_A\rangle = |\Psi_B\rangle = |\Psi(t)\rangle$ and perform 
the sum in (\ref{eq:nx}). We can also obtain two-body qubit correlators,
$\sigma_i^+\sigma_j^-$,  taking $o_{x_i}=\sigma_i^+$, $o_{x_j}= \sigma_j^-$ and $o_n=\mathbb{I}_n$ for $n\neq x_i,x_j$.
Finally, for the projectors $\phi^t_{x_1,\dots,x_N}$,
Eq. (\ref{eq:phiN}), one just takes $|\Psi_A\rangle = |GS\rangle$,
$|\Psi_B\rangle = |\Psi(t)\rangle$, $o_{x_i} = a_{x_i}$ with
$i=1,\dots,N$ and $o_{n} = \mathbb{I}_n$ for $n\neq x_1,\dots,x_N$ .


\section{Linear scattering}
\label{sec:linear}

The main objective of this work is to realise how nonlinear the
scattering of few photons through few qubits is. To accomplish this
task we need first  to know \emph {what linear scattering is}.
In this section we discuss in general what we understand
for linear quantum optics and its realisation in waveguide QED. Finally, we
establish under which conditions a collection of qubits behave as a linear optical medium.


\subsection{Linear systems}
\label{sec:ls}

In quantum physics, \emph{linear systems} are those
whose 
 Heisenberg equations for the observables form a linear set.
For the case of Hamiltonian (\ref{eq:Htot}) this happens whenever the
scatterers are \emph{harmonic resonators} (the waveguide itself is
linear)  both within RWA and non-RWA coupling regimes. In this case,the $c_j,c_j^\dagger$ operators in (\ref{eq:Htot}) are
\emph{bosonic} operators [Cf. Eq. (\ref{c-qubit})]:
\begin{equation}
[c_i, c_j^\dagger] = \delta_{ij}
\end{equation}
and [Cf. Eqs.  (\ref{h-qubit})]:
\begin{equation}
h_i(c_i,c_i^\dagger)=\Delta_ic_i^\dagger c_i \, .
\end{equation}
where $\Delta_i$ are the resonator frequencies.
%


\subsection{Analytical properties for the $S$-matrix whit harmonic resonators as scatterers}
\label{sec:theorem}

Once we know what linear scattering means, we present
our first result.  An
equivalent result was  introduced by us in the Supplementary
Material of~ \cite{Burillo2014}. We re-express it here in a more
convenient way.  
\begin{theorem}
\label{theo1}
If the system is linear (in the sense of Sect. \ref{sec:ls}), the one-photon scattering matrix is given by:
\begin{equation}
\label{eq:S1}
\langle p  | S  | k \rangle
=
t_{k} \delta_{p,k} +r_k \delta_{p,-k}\, .
\end{equation}
\end{theorem}
The apparent simplicity of Theorem \ref{theo1} requires some discussion.
First, photon creation is not possible.  Second, Eq. (\ref{eq:S1}) fixes
the actual form for the output states,
\begin{equation}
\label{eq:eigen1}
| \Psi_{out} \rangle
=
\sum_{k >0}  (t_k \, \phi_k^{in}   a_k^\dagger
+
r_k \, \phi_k^{in} a_{-k}^{\dagger})
| GS \rangle\, .
\end{equation}
Therefore, the only scattering processes for one incoming photon occurring within linear models
are the reflection and transmission of the photon without changing its
input energy (momentum).
Notice that this is a non trivial feature, since the
Hamiltonian (\ref{eq:Htot}) is
not number conserving: $[H, \sum_x a_{x}^\dagger a_{x} + \sum_j c_j^\dagger
c_j] \neq 0$ and 
the ground state $| GS \rangle$ has not got a well defined
number of excitations. 
However, the single photon scatters by reflecting and
transmitting without changing the energy and without creating additional
excitations in the system.  This result
mathematically relies on the Bogolioubov transformation and physically
on the fact that (\ref{eq:Htot}) is a \emph{free} model
in the quantum field theory language .  
The proof of this theorem is sent to Appendix \ref{app:theo1}.

The single photon result, Theorem \ref{theo1}  can be generalised to the
multiphoton 
case:
\begin{theorem}
\label{theo2}
If the system is linear the $N$-photon scattering matrix is given by:
\begin{equation}
\label{eq:Sn}
\langle p_1, p_2, ..., p_N | S | k_1, k_2, ..., k_{N^{\prime}} \rangle
=  
\delta_{N N^\prime}  
\sum_{n_1 \neq n_2 \neq ... \neq n_N}  \langle p_{1}  | S  | k_{n_1} \rangle ... \langle p_{N}  | S  | k_{n_N} \rangle\, .
\end{equation}
\end{theorem}
The theorem says that,
in linear systems the scattering matrix is a product of single
photon scattering matrices.  Consequently, no particle creation,
Raman process or photon-photon interaction is possible.  
The proof, detailed in Appendix \ref{app:theo2}, is based on the single
photon result, Theorem \ref{theo1}, together with the Wick theorem. 
To better appreciate these two general results, let us apply them.


\subsection{The classical limit: Recovering the standard linear optics
  concept}
\label{sec:class}

Theorems \ref{theo1} and
\ref{theo2} dictate the scattering in linear systems (i.e., when the scatterers
are harmonic resonators).  So far, it is not completely clear whether the 
definition for linear systems  in quantum mechanics given in Sect. \ref{sec:ls} together with the
results in section \ref{sec:theorem} correspond to what 
 linear optics means:  the properties for the scattered currents
are independent of the input intensity.
Here we show that linear systems satisfy this intensity independence.
Importantly enough, we  comment on the classical limit for linear
systems.

We consider a monochromatic coherent state as the $N$-photon input state,
\begin{equation}
\label{eq:alpha}
| \Psi_{in} \rangle =
| \alpha_k \rangle = {\rm e}^{-| \alpha_k |^2/2} 
\sum_{n=0}^\infty \frac{\alpha_k^n (a_k^{\dagger})^n}{n!}
|GS \rangle\, .
\end{equation}
Applying the theorems \ref{theo1} and \ref{theo2} the output state can be written as,
\begin{equation}
\label{eq:Psiout}|\Psi_{out}\rangle = 
e^{-|\alpha_k|^2/2}\sum_{n=0}^\infty \frac{\alpha_k^n (t_k a_k^\dagger
  + r_k a_{-k}^\dagger)^n}{n!}|GS\rangle
=
|t_k \alpha_k \rangle \otimes |r_k \alpha_{-k} \rangle \; .
\end{equation}
The second equality follows after some algebra
\footnote{
Just notice that:
\begin{equation*}
\sum_{n=0}^\infty \frac{\alpha_k^n (a_k^{\dagger}(t_{out}))^n}{n!}
=
\sum_{n=0}^\infty \sum_{m=0}^n \frac{1}{m! \, (n-m)!} 
 (t_k \alpha_k a_k^\dagger)^m
(r_k \alpha_k a_{-k}^\dagger)^{n-m}
=
\sum_{n=0}^\infty  \frac{ (t_k \alpha_k a_k^\dagger)^n}{n!}
\sum_{n^\prime=0}^\infty \frac{(r_k \alpha_k a_{-k}^\dagger)^{n^\prime}}{n^\prime!} 
\end{equation*}
}.

Equation (\ref{eq:Psiout}) is a satisfactory result.  
The transmission and reflection coefficients $t_k$ and $r_k$
are independent  of $\alpha_k$.  
Recalling that 
$\langle  \alpha_k | a_k^\dagger a_k | \alpha_k \rangle =
|\alpha_k|^2$, this means independence from
\emph{input intensity}.
We note that 
linear systems
transform coherent  states onto coherent states.  Thus harmonic resonators
do neither change the statistics nor generate  entanglement between
the reflected and transmitted fields.  Coherent states can be
considered classical inputs in the limit $\alpha_k \to \infty$, thus linear systems do
not generate \emph{quantumness}.
Finally,  the last expression for $|\Psi_{out}\rangle$  in (\ref{eq:Psiout}), has the \emph{classical} 
interpretation in terms of transmitted  $|t_k\alpha_k|^2$ and reflected
$|r_k\alpha_k|^2$ currents  ($|t_k|^2 + |r_k|^2 =1$).

\subsection{From nonlinear to linear}
\label{sec:lin-nolin}

Consider $M$ qubits placed at the same point of the waveguide, $x_i =
x_j$ for all $i, j$ in Hamiltonian (\ref{eq:Htot}).  For simplicity,
assume that the couplings $g_i$ are also the same.
Introducing the operator,
\begin{equation}
\label{b-def}
b\equiv \frac{1}{\sqrt{M}}\sum_{i=1}^M \sigma_i^-
\; ,
\end{equation}
the total Hamiltonian, Eq. (\ref{eq:Htot}), can be rewritten
\begin{equation}
H_{\rm tot}=\epsilon\sum_x a_x^\dagger a_x - J \sum_x (a_x^\dagger
a_{x+1} + {\rm  hc} ) 
+ \Delta b^\dagger b + g \sqrt{M}(b^\dagger + b) (a_0^\dagger + a_0)
\; .
\label{eq:Htotb}
\end{equation}
Thus, in terms of the Dicke states generated by (\ref{b-def}) the effective
coupling is $g \sqrt{M}$.  
Besides, it is crucial to observe that 
\cite{Hummer2012}
\begin{equation}
[b, b^\dagger] = 1 - \frac{1}{M} \sum_j \sigma_j^+ \sigma_j^-
\end{equation}
Therefore, in the limit $\langle \sum_j \sigma_j^+ \sigma_j^- \rangle
/ M \ll 1$ (weak probe compared to the number of qubits) the operator
$b$ ($b^\dagger$) approximates an annihilation (creation) bosonic
operator.  Therefore,  a large number of qubits  is expected to behave as a harmonic resonator. 

\section{Nonlinear scattering:  Numerics in the RWA}
\label{sec:results}

We are interested in the nonlinear optical properties of a collection of
qubits.  The
nonlinearities can be manifested in different observables.  The
theoretical results in section \ref{sec:theorem} say nothing about the
nonlinear regime.  
In the following, we will compute some natural quantities as the
reflection and transmission probabilities or photon-photon correlation.  
We will evaluate \emph{how nonlinear
the response is} as a function  of the number of incoming photons, number of
qubits or inter qubit  distance. 
Throughout  this section the RWA is assumed. Physics beyond the RWA is discussed
in the next section.


\subsection{$N$ photons vs $M$ qubits: Total reflection
  spectrum}
\label{sec:RvsNM}

\begin{figure}[t]
\centering
\includegraphics[width=.8\columnwidth]{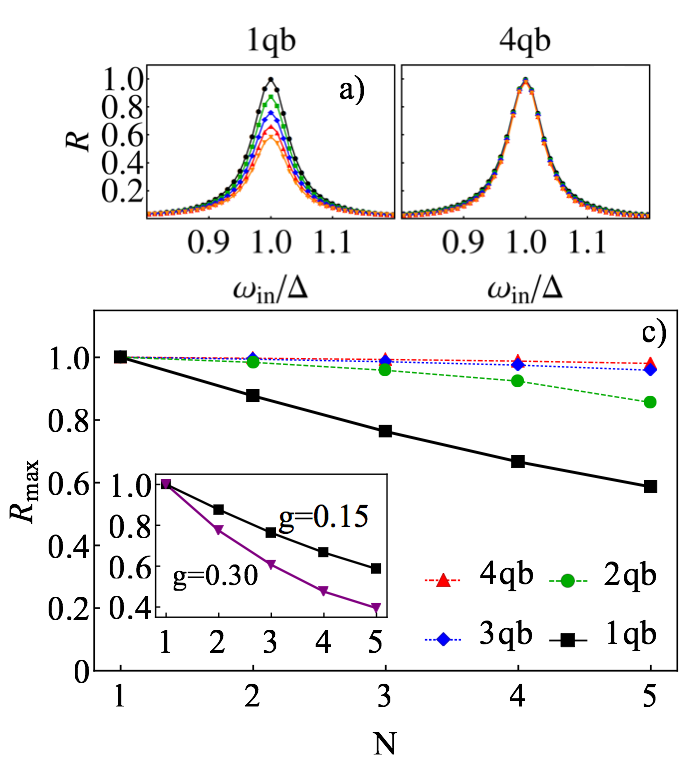}\\
\caption{{\it (Colour online)} $\epsilon$, $J$ and  $\Delta$ as in figure
  \ref{fig:fig_intro}.   The input state, Eqs. (\ref{eq:psi_in}) and
  (\ref{eq:wp}),  with $k_{in}= \pi/2$ and $\theta=2$. a) $R$ for
  $M=1$ and $N=1-5$ (black circles, green squares, blue diamonds, red
  triangles and orange inverted triangles respectively); b) The same
  as in a) but with $M=4$;  c) $R_{max}$ for $g\sqrt{M}=0.15$, with $M=1-5$ qubits (black squares/solid line, green circles/dashed line, blue diamonds/dotted line and red triangles/dotted-dashed line respectively). In the inset we show $R_{max}$ vs $N$ for $M=1$, with $g=0.15$ (black squares) and $g=0.30$ (purple inverted triangles).}\label{fig:ref}
\end{figure}

The combination of energy and number conservation implies that  output states for \emph{one photon}
scattering through qubits in the RWA, are also given by Eq. (\ref{eq:eigen1}) (like in linear
systems).
A well known result in this geometry is that a single monochromatic photon suffers perfect reflection, $|r_{k_{in}}|^2=1$, when its frequency $\omega_{in}=\Delta$\cite{Fan2005a,Fan2005b,Nori2008a}; this effect has several useful applications
\cite{Lukin2007b,Nori2009, Nori2010, Zhou2013, Lu2014}.
For linear systems,  following (\ref{eq:Sn}), the
$N$-photon $S$ matrix is a
product of single photon $S$ matrices.
Thus, in linear systems, 
$N$-photon input
states must also be perfectly reflected at resonance.
On the other hand, a qubit cannot totally reflect 
more than one photon at the same time\cite{Fan2007b}.
Then, for $N$-photon ($N>1$) input states  perfect reflection is not
expected to occur with one qubit.
If we want to overpass this
saturation effect, we may increase the number of qubits.
In the limit $N / M \ll 1$, with $N$ photons and $M$ the number of
qubits the linear regime should be recovered, {\it i.e.}
perfect reflection should happen [See Sect. \ref{sec:lin-nolin}].

Equipped  with the MPS technique we can study the  linear-nonlinear transition as a
function of the ratio $N/M$.  In doing so, we provide meaning to the inequality $N/M \ll 1$.
In this  subsection we  consider that the $M$ qubits are
placed at the same point, i.e., their inter distance is zero. 
 We compute $R_{\omega_{in}}$  from 
$N=1$ to $N=5$ photons input states  given by  (\ref{eq:psi_in})  and (\ref{eq:wp}) centred
in the resonant value $k_{\rm in} = \pi/2$ ($\omega_{k_{in}}=
\Delta$).
The scattering centre is composed by $M=1$ to $M=4$  qubits.  
We plot $R_{\omega_{in}}$
[Cf. Eq. (\ref{eq:RT})]  for  $M=1$ and $M=4$  qubits in  \ref{fig:ref} a) and b) respectively.
The spectral width scales  with the one-photon effective coupling $ g
\sqrt{M}$ (\ref{eq:Htotb}), which is maintained constant in the calculations.
As seen, the maximum reflection 
$R_{max} < 1$ as soon as $N>1$.   The effect is better observed in the
$M=1$ 
qubit  case, see Fig. \ref{fig:ref} a).  As argued before,
by increasing the number of qubits we recover full reflection $R_{max} \cong 1$.
The dependence for $R_{max}$ as a function of $N/M$ is better
represented in panel \ref{fig:ref} c).
$R_{max}$ decreases much faster with $N$ for $M=1$ than for
$M>1$. For $M=4$, $R_{max}$ hardly gets modified by changing the number of
photons in the considered range of $N$. Following\cite{Fan2011a} and\cite{Roy2013}, there is total
reflection for $N=2$ vs $M=2$ if the photon energies individually match
with those of the qubits. We see a slight deviation of
this result because we are taking a non-monochromatic input state and
the component of the incident wavefunction for $k_1,k_2\neq k_{in}$ is
not negligible.
It is remarkable that $R_{max}$ does not only depend on $N$ and $M$,
but also on the coupling, see the inset in  \ref{fig:ref} c).
Therefore the nonlinear characteristics, do not only depend on the
material (qubits) but on their coupling to the field too.


\subsection{Two photons vs $M$ qubits: Spatial photon-photon correlations in reflection}
\label{sec:2vsM}

\begin{figure}[h]
\hspace{0.1cm}\includegraphics[width=1.\columnwidth]{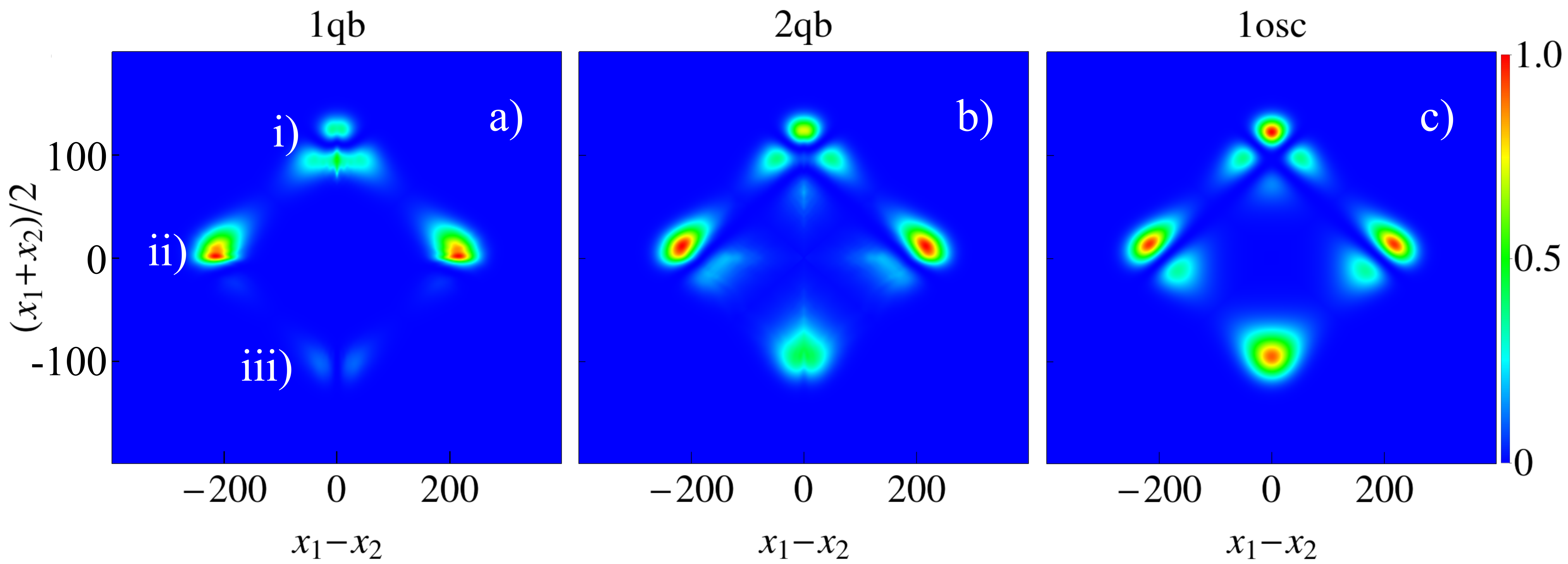}\\
\includegraphics[width=1.0\columnwidth]{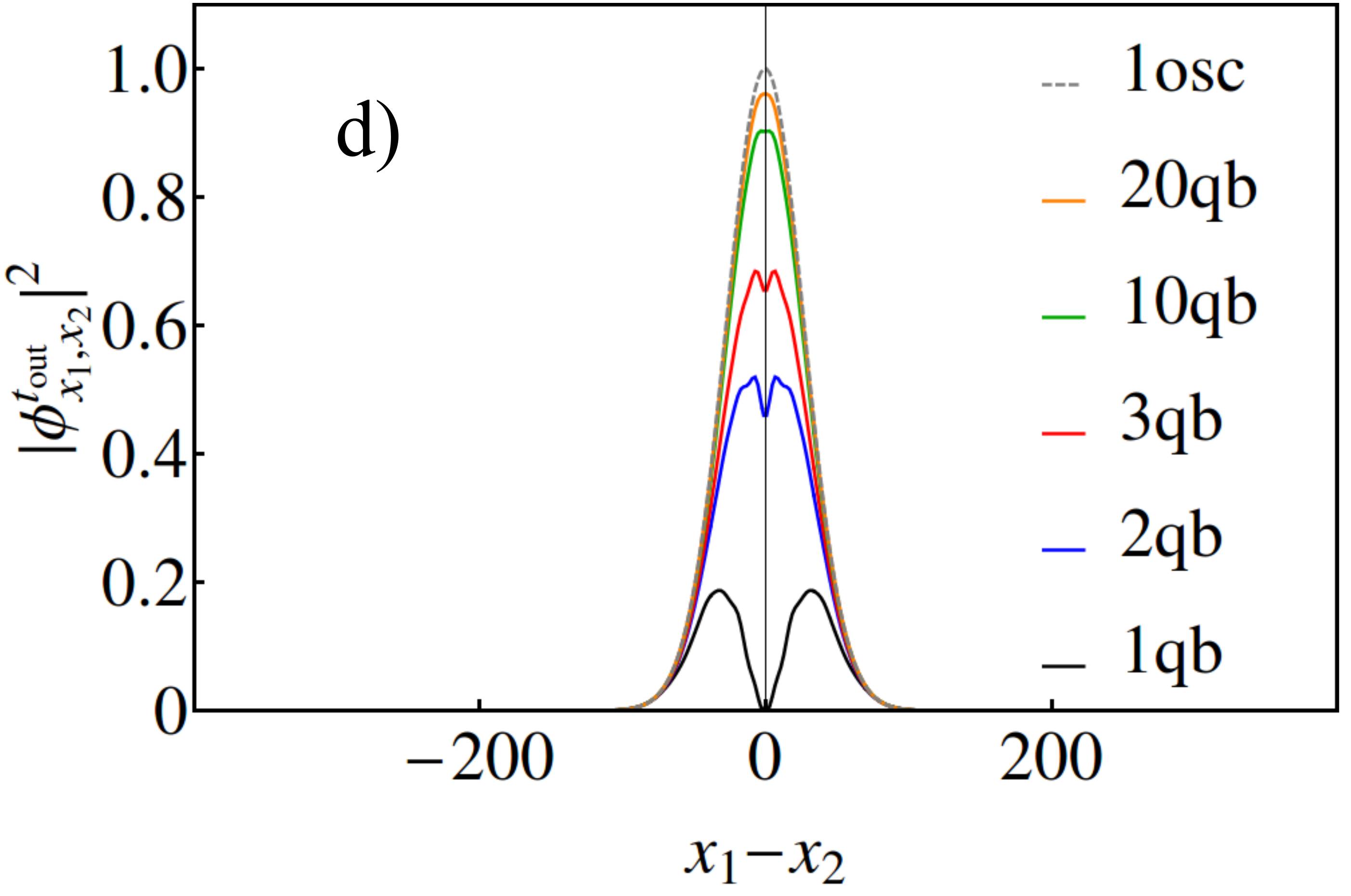}\\
\caption{{\it (Colour online)} 
 $|\phi^{\; t_{out}}_{x_1, x_2}|^2 $
for $g\sqrt{M}=0.10$, with a) $M=1$, b) $M=2$ qubits  and c) a harmonic
oscillator. 
$|\phi^{\; t_{out}}_{x_1, x_2}|^2 $ is normalised in each panel such
that its maximum is set to $1$. In the contours, i) corresponds to the transmitted-transmitted
component, ii) to the transmitted-reflected one and finally iii) to the
reflected-reflected part.
d) 
$|\phi^{\; t_{out}}_{x_1, x_2}|^2 $  fixing $(x_1+x_2)/2$ for $g\sqrt{M}=0.10$, with $M=1,2,3,10,20$ qubits (solid, from
bottom to top) and harmonic oscillator (dashed).
 We normalise such that  correlator for the harmonic oscillator is  $1$ for $x_1=x_2$. 
The input state is a $2$-photon state with
$k_{in}=\pi/2$ and $\theta=20$.
}
\label{fig:antibunching1}
\end{figure}

In this section we compute the photon-photon correlations created in the scattering process for two-photon input states, as a function of the number of qubits.  
%
%
For further comparison and understanding, we begin by discussing the
linear case.  Then the scatterer is a harmonic oscillator
  $h = \Delta
c^\dagger c$ with $[c, c^\dagger]= 1$ in Eq. (\ref{eq:Htot}).
It has already been discussed in this paper that  no correlations are
generated if the scatterer is linear [Cf. Theo.
\ref{theo2}].
The two point correlation factorises:
 $|\phi^{\; t_{out}}_{x_1, x_2}|^2 = |\phi^{\;
  t_{out}}_{x_1}|^2 |\phi^{\; t_{out}}_{x_2}|^2$ and, 
in particular, the two photons must be bunched both in reflection and transmission:  
$|\phi^{\; t_{out}}_{x, x}|^2 = |\phi^{\;
  t_{out}}_{x}|^4$.
In the nonlinear case, however,  the reflected field by one qubit is totally antibunched  \cite{Fan2007b}, 
$|\phi^{\; t_{out}}_{x, x}|^2=0$.  
Thus, antibunching can be used as 
witness for nonlinearities. 
With these antecedents, we provide below answers to the
following questions.
How does this depend on the number of qubits?  Is it possible to
interpolate between the highly nonlinear case of one qubit and the
linear case of a harmonic oscillator by adding qubits? If so, how many qubits are needed for the system to be linear?

In figure
\ref{fig:antibunching1}~d) we plot the reflected
$|\phi_{x_1,x_2}|^2$  against
 $x_1-x_2$, fixing $(x_1+x_2)/2$ such that the reflection component is
 maximum. 
The one-photon coupling, $g\sqrt{M}$ is kept constant for different $M$
and equals the coupling
$g$ for the case of the harmonic oscillator.
We remind that $|\phi_{x_1,x_2}|^2$
is proportional to the probability of having both
photons 
separated by a distance $|x_1-x_2|$.
Setting $x_1=x_2$ gives the probability of seeing both photons at the
same point. 
For  $M=1$  the numerical results (Fig. \ref{fig:antibunching1} d)
show $|\phi_{xx}|^2\simeq 0$, recovering the well known photon
antibunching in reflection \cite{Fan2007b,Fan2011a}. Surprisingly, photon
antibunching and thus nonlinearity can still be resolved by increasing $M$, even for $M=20$
\footnote{We note here that for the particular set of values $N=2$ and
  $M=2$ that, if $\omega_1=\Delta_1$ and $\omega_2=\Delta_2$ photons
  are bunched \cite{Fan2011a,Roy2013}as it would be linear scattering.  However, our initial
  wavepacket is not monochromatic }.
The full contour for $|\phi_{x_1,x_2}|^2$ is shown in
\ref{fig:antibunching1} a), b) and c) for one, two qubits and harmonic
resonator respectively.
Apart from the antibunched characteristics in the reflected signal, we
can also appreciate that one qubit cannot reflect as much as two
qubits, as we discussed in Sect. \ref{sec:RvsNM}
[Cf. Fig. \ref{fig:ref}].
 The
transmitted photons are always bunched. 
The components around $(x_1+x_2)/2\simeq 0$ and $|x_1-x_2|\simeq 200$
correspond to one photon being transmitted and the other reflected.

Once the physics has been discussed, let us finish with a brief note
on how to solve  the two photon scattering against $M$
qubits, for any $M$, placed at the same point and within the RWA.
We start by reminding that the RWA implies the conservation for the number of
excitations,
Cf. Sect \ref{sec:physical}.

Therefore, in the two excitation manifold and regarding the qubits, it will suffice to consider the following
Dicke states: 
$ \{ |0\rangle \, , |1\rangle \equiv b^\dagger|0\rangle \, , |2\rangle \equiv
\frac{(b^\dagger)^2|0\rangle}{\sqrt{2(1-1/M)}} \} $ [Cf. Eq. (\ref{b-def})].
Thus, if $N=2$, the $M$ qubits  can be
formally replaced by  a three level system with states given as above.
As expected [See Sect. \ref{sec:lin-nolin}], in the limit $M\to \infty$,  $|2\rangle =
\frac{(b^\dagger)^2|0\rangle}{\sqrt{2}}$ as in the harmonic
oscillator case.


\subsection{Three qubits with distance}
\label{sec:2ph3qb}

\begin{figure}[h]
\begin{center}
\includegraphics[width=1.\columnwidth]{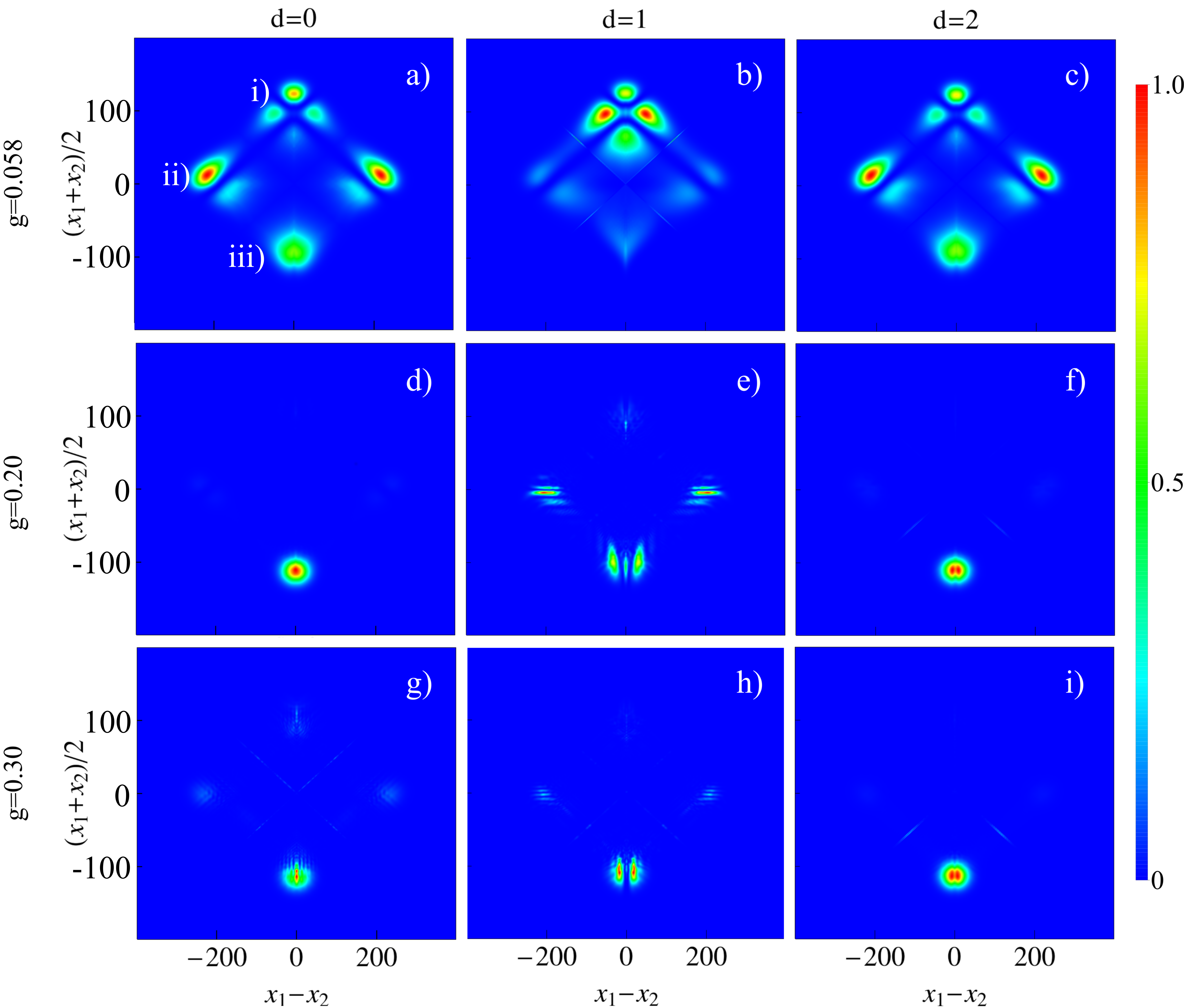}\\
\caption{{\it (Colour online)}
 $|\phi_{x_1,x_2}^{t_{out}}|^2$ for two incident photons and three
qubits with inter-qubit distance $d$. i), ii) and iii) mean the same as in Fig. \ref{fig:antibunching1}. The incoming photons are characterised by $\theta=20$ and
$k_{in} = \pi/2$. We take $d=0,1,2$, (left, middle and right columns respectively), 
so the first and the third cases
should be equivalent because of (\ref{eq:condition}). We consider the
RWA Hamiltonian and take $g\sqrt{3}=0.10$ in the first row, whereas we take the full Hamiltonian
for the second and third rows, with $g=0.20$ and $g=0.30$ respectively.}
\label{fig:phi2_3qb}
\end{center}
\end{figure}

\begin{figure}[h]
\begin{center}
\includegraphics[width=1.\columnwidth]{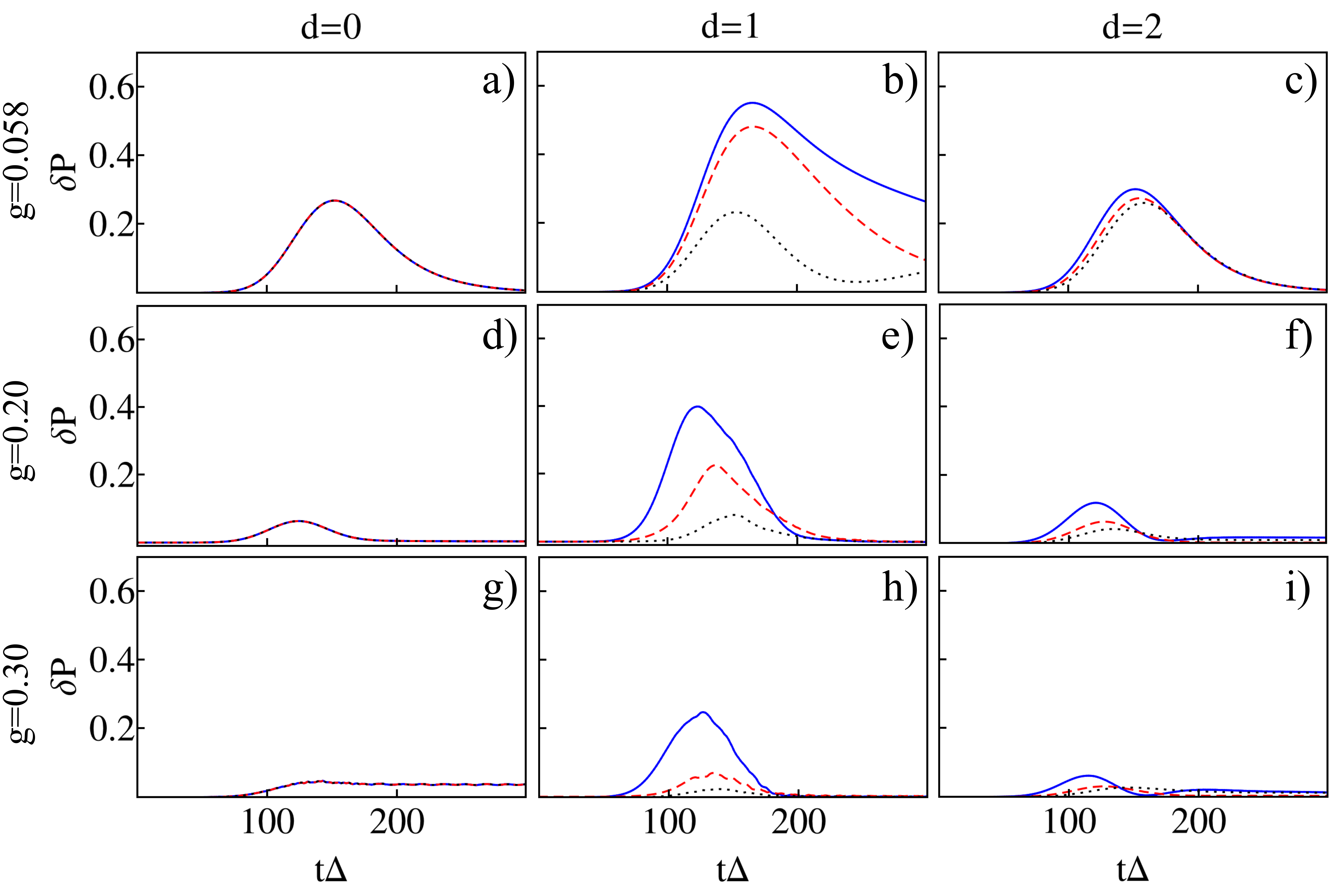}\\
\caption{{\it (Colour online)} Qubit population of the first (solid blue), the second (dashed red) and the third qubit (dotted black). The parameters are those of figure \ref{fig:phi2_3qb}.}
\label{fig:Pqb_3qb}
\end{center}
\end{figure}

We incorporate a new ingredient here,  the inter-qubit
distance.
Interactions among qubits mediated by the EM field
decay with the distance in two and three dimensions \cite{Lehmberg1970}.  
One dimension
is special: the field wavefront area does not grow with
distance and, thus,  the interaction does not decay but it is
periodically modulated instead.
The period depends on the qubits level splitting and the dispersion relation
in the waveguide.
This peculiarity has been theoretically investigated for
qubits in a multitude  of arrangements
\cite{Nori2008b, Tudela2011a,  Zueco2012, Wallraff2013a, Ballestero2014} and
experimentally demonstrated quite recently \cite{Wallraff2013b}.
For our purposes it is important to note that these works considered
either classical or single photon input states.
Two photon input states were considered too in
\cite{Baranger2013,Baranger2014}.
%

Invoking the  Markovian approximation the induced
qubit-qubit interaction is considered instantaneous. This implies that the set of distances, $d\equiv
x_{i^\prime} -x_i$, related by 
\begin{equation}
\label{eq:condition} 
d^\prime= d+  \frac{ \pi }{k_{in}} q,\qquad q\in\mathbb{Z}
\end{equation}
give the same scattering characteristics. 

Let us consider $N=2$ input photon states and three qubits separated
by some nearest neighbour distance $d$.
With the MPS tool,   we solve the problem exactly, no matter the
distance. Thus, we do not make any approximation like the small distance or Markovian ones.
We plot photon-photon correlations $|\phi_{x_1,x_2}^{t_{out}}|^2$ (\ref{eq:phiN})  and qubit populations
$\delta P_i\equiv P_i - (P_i)_{GS}$ (\ref{eq:qubit}) for \emph{RWA couplings} ($(P_i)_{GS}=0$) 
in figures  \ref{fig:phi2_3qb} and \ref{fig:Pqb_3qb}, panels  a), b) and c) in
both cases.
Regarding the photons, the characteristics are the same
at zero distance and 
$d=2$, which is equivalent to zero distance, by means of (\ref{eq:condition})
($k_{in} = \pi/2$). 
On the other hand, the inherent non-Markovian properties of our exact simulation
can be appreciated by comparing  \ref{fig:Pqb_3qb} a) and c).
If $d\neq 0$, the qubit at the left is excited first, then the central
qubit and finally the one to the right.

The contours for the two-photon wave function 
$|\phi_{x_1,x_2}^{t_{out}}|^2$ are for $d=0$ and $d=2$ closer to the linear scattering
result than those for $d=1$ case [Cf. figures
\ref{fig:antibunching1} c)  and \ref{fig:phi2_3qb} a) b) and c)].
This enhancement of nonlinear properties with the distance can be understood as follows.
At distances $ \frac{ \pi }{k_{in}} q$, [equivalent to zero distance according to
Eq. (\ref{eq:condition})], only  qubits states generated by the ladder
operator $b$ in Eq.  (\ref{b-def}) are visited  during the dynamics.
For  two photon input states, the  nonlinearities
die out as  $1/M$ [Cf. Sect. \ref{sec:2vsM}].
On the other hand, for other distances not fulfilling this condition
of equivalence, more qubit states (satisfying the number
conservation imposed by the RWA) can play a role.  
The fact that more qubit modes participate in the dynamics for $d=1$
is apparent from 
 figure \ref{fig:Pqb_3qb} b).   Clearly, more frequencies are involved
 in the evolution of 
$P$.
Importantly enough, $d=1$ corresponds to a distance where the coherent
interaction between the qubits are maximised, while the correlations
in the qubit decays are minimised.  This is named as subradiant
case.  As we can observe the qubit decay (after excitation) is slower
in this case.
%


\section{Nonlinear scattering in the ultrastrong}
\label{sec:ultrastrong}

We close this \emph{Discussions} by investigating stronger
qubit-photon couplings, strong enough to
break down the RWA approximation,  Eq. (\ref{eq:hintrwa}).  The 
 full dipole
interaction must be considered.  
When moving to the ultrastrong regime, the problem becomes rather involved.
The ground state, $|GS\rangle$, for (\ref{eq:Htot}) is not trivial anymore but strongly correlated and the
number of excitations is not conserved [Recall
Sect. \ref{sec:physical}].
This increase in the complexity brings a lot of new phenomena.  
We fix our attention in two characteristics. First, we discuss  the
appearance of 
inelastic scattering  at high
couplings \cite{Burillo2014, Longo2010, Longo2011,Sun2009b}.  Second,  we
revisit the qubit-qubit interactions mediated by the guide   in the ultrastrong
coupling regime.

\subsection{Inelastic scattering with $M$ qubits}
\label{sec:Raman}

\begin{figure}[h]
\centering
\includegraphics[scale=.4]{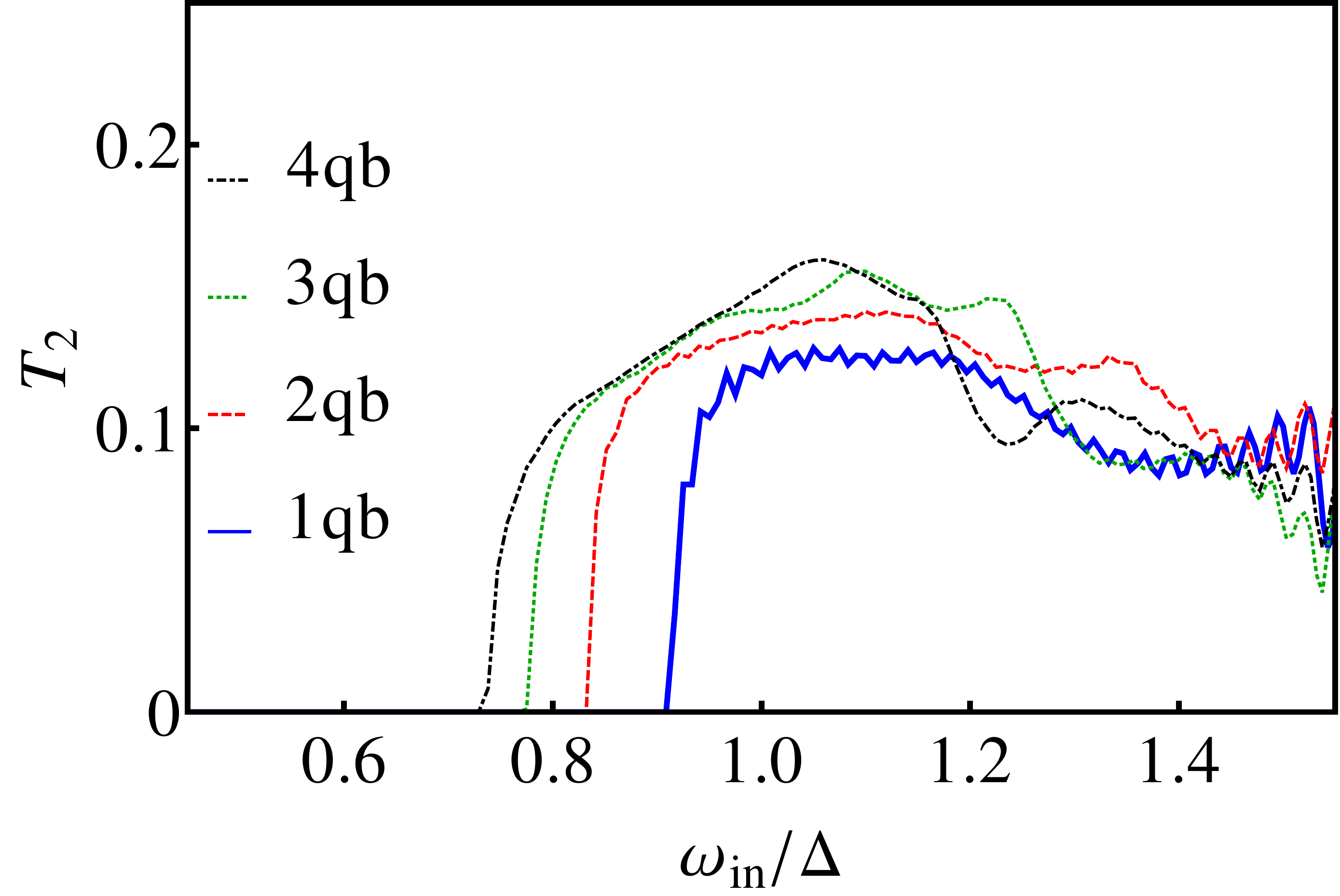}
\caption{{\it (Colour online)} Inelastic transmission probability for $g\sqrt{M}=0.50$ and 1 qubit (solid blue), 2 qubits (dashed red), 3 qubits (dotted green) and 4 qubits (dashed-dotted black).}
\label{fig:raman}
\end{figure}

So far, we have discussed elastic scattering.  In linear systems, this is the only possible process.  When the scatterers are qubits, however,
inelastic processes may happen \cite{Burillo2014,Longo2010, Longo2011,Sun2009b}. 
In an inelastic process, the photons can excite localised
light-matter states.  Those localised states, eigenstates for the
total Hamiltonian, remain excited after the photons pass through.  By energy
conservation it is clear that the outgoing photons must have less
energy.
Within the RWA approximation\cite{Longo2010, Longo2011,Sun2009b}, at least two photons are needed
for having inelastic processes.
When the actual full dipole interaction is considered
(\ref{eq:Htot}), non stimulated Raman scattering process occurs \cite{Burillo2014}.
Indeed, it is quite remarkable that $100 \%$ efficiency can be
achieved \cite{Burillo2014}. 

We study the inelastic channel for one photon input states when
interacts with $M$ qubits placed at the same point.  In figure \ref{fig:raman} we plot the
transmitted current in the inelastic channel,
\begin{equation}
\label{T2}
T_{2,k} = \frac{1}{2}(1 - |t_k|^2 - |r_k|^2)
\end{equation}
In the figure we consider the qubits placed in the same point.
This figure shows that inelastic scattering is another nonlinear characteristic
which persists even for $N \ll M$ (recall that
$T_2$ for a linear system is always zero, Cf. Sect. \ref{theo1}).

\subsection{Mixing distance and ultrastrong}
\label{sec:distance_ultrastrong}

Interesting physics occurs with spatially separated
qubits, when their interactions are mediated through the photons.  At the same
time the photons  interact among themselves when they pass through
those qubits.  In the weak coupling regime, the output field is \emph{periodic}
in the qubit-qubit distance when taking Markovian approximation (\ref{eq:condition})  [See section
\ref{sec:2ph3qb}].
Below, we discuss how this scenario changes in the ultrastrong coupling
regime.

In figure  \ref{fig:phi2_3qb} the loss of \emph{this effective
  periodicity}  in the qubit distance is clearly shown by plotting
$|\phi_{x_1,x_2}^{t_{out}}|^2$.
At weak coupling (first row, already discussed in Sect
\ref{sec:2ph3qb}) the distances $d=0$ and $d=2$ satisfying
(\ref{eq:condition})  present identical  output fields. 
The two lower rows show results at larger 
couplings.  The condition (\ref{eq:condition}) does not hold anymore,
and the dipole-dipole periodic structure paradigm  for one dimensional
systems \cite{Nori2008b, Tudela2011a,  Zueco2012, Wallraff2013a,
  Wallraff2013b}  is not longer true.
Further confirmation is obtained when looking to the qubit dynamics.
In the final column, we realise that,  in the ultrastrong regime,
signatures of subradiant dynamics are still perceived.  The loss of
periodicity is also appreciated here.  If we fix our attention
to the  $d=0$ case all qubits behave in the same manner, as they must,
and they remain in an excited state, which is a
signature of inelastic scattering. For the case $d=2$, which in weak
coupling is equivalent to $d=0$ [Cf. Eq. (\ref{eq:condition})],  the dynamics is completely different.
 At this set
of parameters the qubits seem to evolve back to the ground state.
Therefore no inelastic scattering occurs in this case.

\section{Conclusions}

The advent of artificial devices (optical cavities,
superconducting circuits, dielectric and plasmonic guides,
photonic crystals)  opens the avenue for stronger and stronger light-matter
coupling (at the single photon level).
Recent impressive experimental advances are changing the paradigm
and typical nonlinear effects can be observed at the few photon level. 
From the theoretical point of view, both light and matter must be
described quantum mechanically. 
Going beyond one or two photons
and one qubit is an analytic titanic task. 
Therefore, numerical tools are demanded to satisfy current experimental
efforts in constructing  devices responding nonlinearly at minimum
powers.
In this \emph{Discussions} we have shown that the MPS technique
developed for one dimensional systems is a powerful tool in few photon
nonlinear optics.

On the physical side, our task was to study the nonlinear
response by increasing the number of qubits.  It has a practical
motif.  The effective coupling scales with $g
\sqrt{M}$ ($M$ the number of qubits).  Adding scatterers is a way to
enhance the coupling but, at the same time,  their nonlinear
response is reduced.
Some quantities, as the transmitted and reflected
currents, already gave a linear behaviour for ratios  $N / M \cong 1$.
However, we have found that for $N = 2$
input states, $M=20$ qubits  still generate  photon-photon
interactions.

We have also investigated the regime where  light and matter are
coupled ultrastrongly. New phenomena appears, as Raman scattering or the 
breakdown of the periodicity in the qubit-qubit interaction, which should
be obesrvable by current technology, since some experimental setups already
operate in the ultrastrong coupling regime.
All together, serves as a motivation for  developing theories for its
understanding and hopefully, will trigger further experimental studies for verifying the plethora of new
phenomenology.

\section*{Acknowledgements}

We acknowledge 
support by the Spanish Ministerio de Economia y Competitividad within projects MAT2011-28581-C02, FIS2012-33022 and No. FIS2011-25167, the Gobierno
de Aragon (FENOL group)
and the European project PROMISCE.

\appendix

\section{Proof of theorem \ref{theo1}}
\label{app:theo1}

If the scatterers are harmonic resonators the
Hamiltonian (\ref{eq:Htot}) is linear [Cf. Sect. \ref{sec:ls}]  and it can be diagonalised with a
Bogolioubov-Valatin (BV) transformation,
\begin{equation}
\label{eq:H-normal}
H = \sum \Lambda_l \alpha^\dagger _l \alpha_l \, ,
\end{equation}
with $[\alpha_l,
\alpha_m^\dagger]=\delta_{lm}$ and $\Lambda_l > 0$ $\forall \,
l$.  The ground state is
$\alpha_l | G S \rangle = 0$ $\forall \; l$.
The BV transformation 
\begin{equation}
\label{eq:alpha-a}
\alpha_l = \sum_{i=-L}^L (\chi_{li}^a a_i + \eta_{li}^a a_i^\dagger) + \sum_{i=1}^M (\chi_{li}^c c_i + \eta_{li}^c c_i^\dagger).
\end{equation}
is invertible.

Hamiltonian (\ref{eq:H-normal}) commutes with
$N_\alpha = \sum \alpha_l^\dagger \alpha_l $:
\begin{equation}
\label{HN}
[H, N_\alpha]=0 \; .
\end{equation}
The number of $\alpha$-excitations, $N_\alpha$, is a good quantum number.
It also conserves parity, $P= {\rm e}^{i \pi \sum
  \alpha_l^\dagger \alpha_l }$:
\begin{equation}
\label{HP}
[H, P]=0\; .
\end{equation}


Single photon input states, $N=1$ in  (\ref{eq:psi_in}),  are written in momentum space
\begin{equation}
\label{eq:psi_in_k}
| \Psi_{in} \rangle
= 
\sum_{k>0} \tilde{\phi}_k^{in} a_k^\dagger | GS \rangle,
\end{equation}
with $\tilde{\phi}_k^{in}$ the Fourier transform of $\phi_x^{in}$.
Using the  transformation (\ref{eq:alpha-a}) and that
$\alpha_l|GS\rangle=0$, we can rewrite the  state (\ref{eq:psi_in_k}) in the $\alpha$-representation 
\begin{equation}
\label{eq:psi_in_alpha}
| \Psi_{in} \rangle
= 
\sum_l \bar{\phi}_l^{in} \alpha_l^\dagger  |GS\rangle.
\end{equation}

Given the input state (\ref{eq:psi_in_alpha}):  $N_\alpha | \Psi_{in}
\rangle = | \Psi_{in}
\rangle $. Since $N_\alpha$ is a conserved quantity [Cf. Eq. (\ref{HN})]  the time evolution is restricted to the {\it one $\alpha$-excitation
  level} (\ref{HN}). The output state is then
\begin{equation}
|\Psi_{out}\rangle = \sum_l  \bar{\phi}_l^{out} \alpha_l^\dagger|GS\rangle,
\end{equation}
with $\bar{\phi}_l^{out} \equiv e^{-i\Lambda_l
  t_{out}}\bar{\phi}_l^{in}$. 
Using the transformation (\ref{eq:alpha-a}),  the output state is rewritten
\begin{equation}
\label{eq:psi_out_1}| \Psi_{out} \rangle =
\sum_{l} \bar{\phi}_l^{out}\left(\sum_k((\tilde{\chi}_{lk}^a)^* a_k^\dagger + (\tilde{\eta}_{lk}^a)^* a_k )
+\sum_i((\chi_{li}^c)^* c_i^\dagger +(\eta_{li}^c)^* c_i)\right) | GS \rangle,
\end{equation}
with $\tilde{\chi}_{lk}^a$ and $\tilde{\eta}_{lk}^a$ the discrete
Fourier transforms of $\chi_{li}^a$ and $\eta_{li}^a$ in the second
index.  The output state
(\ref{eq:psi_out_1}) removes the possibility of having multiphoton
scattering states.
Therefore, the scattering events can be elastic,
with transmission and reflection amplitudes $t_{k}$ and $r_{k}$
and inelastic, with the scatterer relaxing to
an excited state $|EXC\rangle$. In the latter, the photon emerges with a new momentum $k_{new}$, fulfilling energy conservation
\begin{equation}
\label{Eine}
\omega_{k_{in}} + E_{GS} = \omega_{k_{new}} + E_{EXC}
\end{equation}
Therefore, the output state can be rewritten 
\begin{equation}
\label{eq:psi_out_2}|\Psi_{out}\rangle = \sum_{k>0} \tilde{\phi}_k^{in}(t_k a_k^\dagger + r_k a_{-k}^\dagger)|GS\rangle + \sum_k \tilde{\phi}_k^{new} a_k^\dagger|EXC\rangle,
\end{equation}
with $\tilde{\phi}_k^{new}$ a wavepacket centred around $k_{new}$. 

Let us fix our attention to the second term in the r.h.s of
(\ref{eq:psi_out_2}), which is rewritten in terms of the 
$\alpha$-operators with the help of the BV transformation (\ref{eq:alpha-a}):
\begin{equation}
\label{eq:raman_term}\sum_k \tilde{\phi}_k^{new} a_k^\dagger|EXC\rangle 
 = \sum_l(\bar{\phi}_l^{new,p} \alpha_l^\dagger + \bar{\phi}_l^{new,m} \alpha_l)|EXC\rangle
\end{equation}
Using $N_\alpha$ and $P$ conservation, Eqs. (\ref{HN}) and (\ref{HP}), 
$N_\alpha |EXC\rangle = 2n|EXC\rangle$ with $n\geq 1$. The first term
in the r.h.s. of (\ref{eq:raman_term}) has $2n+1\geq 3$ particles.  
Thus, $\bar{\phi}_l^{new,p}=0$.  Finally 
$\alpha_l|EXC\rangle$ is  an eigenstate of  (\ref{eq:H-normal}) with
eigen-energy  $E_{EXC}-\Lambda_l$ ($\Lambda_l >0$).  The latter must equal to
$\omega_{k_{new}} + E_{EXC}$ which is impossible.  Therefore $
\bar{\phi}_l^{new,m}=0$.  

Putting all together, the output state contains only the  elastic channel,
\begin{equation}
\label{eq:psi_out_f}|\Psi_{out}\rangle = \sum_{k>0} \tilde{\phi}_k^{in}(t_k a_k^\dagger + r_k a_{-k}^\dagger)|GS\rangle.
\end{equation}
This ends the proof.

\section{Proof or theorem \ref{theo2}}
\label{app:theo2}

The components for the scattering matrix (\ref{eq:Spk}) can be
rewritten,
 \begin{equation}
S_{p_1  ...  p_{N^{\prime}} \, , \, k_1 ...  k_N}
=
\langle GS  | a_{p_1} ... a_{p_{N^\prime}}    \, a_{k_1}^\dagger (t_{out})
... a_{k_{N}}^\dagger(t_{out})  | GS \rangle
\end{equation}
with, 
\begin{equation}
\label{eq:akd_out}a_k^\dagger (t_{out}) = S^\dagger a_k^\dagger S = (t_k a_k^\dagger + r_k a_{-k}^\dagger)
\, .
\end{equation}
In the last equality we have used \emph{i)} linearity:  the Heisenberg
evolution for the operators $a_{k_n}(t)$ is independent of the input
states and \emph{ii)} the theorem \ref{theo2}.
Equation
 (\ref{eq:akd_out}), together with the   Wick theorem:
\begin{equation}
\langle GS | a_{p_1} ... a_{p_{N^\prime}}    \, a_{k_1}^\dagger 
... a_{k_{N}}^\dagger| GS \rangle
= \delta_{NN^\prime}
\sum_{m_1 \neq m_2 \neq ... \neq m_N}
\delta_{p_1 \, k_{m_1}}   ... \delta_{p_N \, k_{m_N}}
\end{equation}
completes the proof.

\footnotesize{
\bibliography{scattering_eduardo} 

\providecommand*{\mcitethebibliography}{\thebibliography}
\csname @ifundefined\endcsname{endmcitethebibliography}
{\let\endmcitethebibliography\endthebibliography}{}
\begin{mcitethebibliography}{72}
\providecommand*{\natexlab}[1]{#1}
\providecommand*{\mciteSetBstSublistMode}[1]{}
\providecommand*{\mciteSetBstMaxWidthForm}[2]{}
\providecommand*{\mciteBstWouldAddEndPuncttrue}
  {\def\EndOfBibitem{\unskip.}}
\providecommand*{\mciteBstWouldAddEndPunctfalse}
  {\let\EndOfBibitem\relax}
\providecommand*{\mciteSetBstMidEndSepPunct}[3]{}
\providecommand*{\mciteSetBstSublistLabelBeginEnd}[3]{}
\providecommand*{\EndOfBibitem}{}
\mciteSetBstSublistMode{f}
\mciteSetBstMaxWidthForm{subitem}
{(\emph{\alph{mcitesubitemcount}})}
\mciteSetBstSublistLabelBeginEnd{\mcitemaxwidthsubitemform\space}
{\relax}{\relax}

\bibitem[Boyd(2003)]{Boyd2003}
R.~W. Boyd, \emph{{Nonlinear Optics, Second Edition}}, Academic Press, 2nd edn,
  2003\relax
\mciteBstWouldAddEndPuncttrue
\mciteSetBstMidEndSepPunct{\mcitedefaultmidpunct}
{\mcitedefaultendpunct}{\mcitedefaultseppunct}\relax
\EndOfBibitem
\bibitem[Peyronel \emph{et~al.}(2012)Peyronel, Firstenberg, Liang, Hofferberth,
  Gorshkov, Pohl, Lukin, and Vuletic]{Lukin2012b}
T.~Peyronel, O.~Firstenberg, Q.-Y. Liang, S.~Hofferberth, A.~V. Gorshkov,
  T.~Pohl, M.~D. Lukin and V.~Vuletic, \emph{Nature}, 2012, \textbf{488},
  57\relax
\mciteBstWouldAddEndPuncttrue
\mciteSetBstMidEndSepPunct{\mcitedefaultmidpunct}
{\mcitedefaultendpunct}{\mcitedefaultseppunct}\relax
\EndOfBibitem
\bibitem[Firstenberg \emph{et~al.}(2013)Firstenberg, Peyronel, Liang, Gorshkov,
  Lukin, and Vuletić]{Lukin2013}
O.~Firstenberg, T.~Peyronel, Q.-Y. Liang, A.~V. Gorshkov, M.~D. Lukin and
  V.~Vuletić, \emph{Nature}, 2013, \textbf{502}, 71\relax
\mciteBstWouldAddEndPuncttrue
\mciteSetBstMidEndSepPunct{\mcitedefaultmidpunct}
{\mcitedefaultendpunct}{\mcitedefaultseppunct}\relax
\EndOfBibitem
\bibitem[You and Nori(2011)]{You2011}
J.~Q. You and F.~Nori, \emph{Nature}, 2011, \textbf{474}, 589--97\relax
\mciteBstWouldAddEndPuncttrue
\mciteSetBstMidEndSepPunct{\mcitedefaultmidpunct}
{\mcitedefaultendpunct}{\mcitedefaultseppunct}\relax
\EndOfBibitem
\bibitem[Astafiev \emph{et~al.}(2010)Astafiev, Zagoskin, Abdumalikov, Pashkin,
  Yamamoto, Inomata, Nakamura, and Tsai]{Astafiev2010}
O.~Astafiev, A.~M. Zagoskin, A.~A. Abdumalikov, Y.~A. Pashkin, T.~Yamamoto,
  K.~Inomata, Y.~Nakamura and J.~S. Tsai, \emph{Science}, 2010, \textbf{327},
  840\relax
\mciteBstWouldAddEndPuncttrue
\mciteSetBstMidEndSepPunct{\mcitedefaultmidpunct}
{\mcitedefaultendpunct}{\mcitedefaultseppunct}\relax
\EndOfBibitem
\bibitem[Hoi \emph{et~al.}(2013)Hoi, Wilson, Johansson, Lindkvist, Peropadre,
  Palomaki, and Delsing]{Johansson2013b}
I.~C. Hoi, C.~M. Wilson, G.~Johansson, J.~Lindkvist, B.~Peropadre, T.~Palomaki
  and P.~Delsing, \emph{New Journal of Physics}, 2013, \textbf{15},
  025011\relax
\mciteBstWouldAddEndPuncttrue
\mciteSetBstMidEndSepPunct{\mcitedefaultmidpunct}
{\mcitedefaultendpunct}{\mcitedefaultseppunct}\relax
\EndOfBibitem
\bibitem[van Loo \emph{et~al.}(2013)van Loo, Fedorov, Lalumi\`{e}re, Sanders,
  Blais, and Wallraff]{Wallraff2013b}
A.~F. van Loo, A.~Fedorov, K.~Lalumi\`{e}re, B.~C. Sanders, A.~Blais and
  A.~Wallraff, \emph{Science}, 2013, \textbf{342}, 1494\relax
\mciteBstWouldAddEndPuncttrue
\mciteSetBstMidEndSepPunct{\mcitedefaultmidpunct}
{\mcitedefaultendpunct}{\mcitedefaultseppunct}\relax
\EndOfBibitem
\bibitem[Lund-Hansen \emph{et~al.}(2008)Lund-Hansen, Stobbe, Julsgaard,
  Thyrrestrup, S\"unner, Kamp, Forchel, and Lodahl]{Lodahl2008}
T.~Lund-Hansen, S.~Stobbe, B.~Julsgaard, H.~Thyrrestrup, T.~S\"unner, M.~Kamp,
  A.~Forchel and P.~Lodahl, \emph{Physical Review Letters}, 2008, \textbf{101},
  113903\relax
\mciteBstWouldAddEndPuncttrue
\mciteSetBstMidEndSepPunct{\mcitedefaultmidpunct}
{\mcitedefaultendpunct}{\mcitedefaultseppunct}\relax
\EndOfBibitem
\bibitem[O'Brien(2007)]{OBrien2007}
J.~L. O'Brien, \emph{Science}, 2007, \textbf{318}, 1567\relax
\mciteBstWouldAddEndPuncttrue
\mciteSetBstMidEndSepPunct{\mcitedefaultmidpunct}
{\mcitedefaultendpunct}{\mcitedefaultseppunct}\relax
\EndOfBibitem
\bibitem[Rezus \emph{et~al.}(2012)Rezus, Walt, Lettow, Renn, Zumofen, Gotzinge,
  and Sandoghdar]{Sandoghdar2012}
Y.~Rezus, S.~Walt, R.~Lettow, A.~Renn, G.~Zumofen, S.~Gotzinge and
  V.~Sandoghdar, \emph{Physical Review Letters}, 2012, \textbf{108},
  093601\relax
\mciteBstWouldAddEndPuncttrue
\mciteSetBstMidEndSepPunct{\mcitedefaultmidpunct}
{\mcitedefaultendpunct}{\mcitedefaultseppunct}\relax
\EndOfBibitem
\bibitem[Boyd(2007)]{Maier2007}
R.~W. Boyd, \emph{{Plasmonics: Fundamentals and Applications}}, Springer, 2nd
  edn, 2007\relax
\mciteBstWouldAddEndPuncttrue
\mciteSetBstMidEndSepPunct{\mcitedefaultmidpunct}
{\mcitedefaultendpunct}{\mcitedefaultseppunct}\relax
\EndOfBibitem
\bibitem[Chang \emph{et~al.}(2006)Chang, S., R., and Lukin]{Lukin2006}
D.~E. Chang, S.~A. S., H.~P. R. and M.~Lukin, \emph{Physical Review Letters},
  2006, \textbf{97}, 053002\relax
\mciteBstWouldAddEndPuncttrue
\mciteSetBstMidEndSepPunct{\mcitedefaultmidpunct}
{\mcitedefaultendpunct}{\mcitedefaultseppunct}\relax
\EndOfBibitem
\bibitem[Akimov \emph{et~al.}(2007)Akimov, Mukherjee, Yu, Chang, Zibrov,
  Hemmer, Park, and Lukin]{Lukin2007b}
A.~V. Akimov, A.~Mukherjee, C.~L. Yu, D.~E. Chang, A.~S. Zibrov, P.~R. Hemmer,
  H.~Park and M.~D. Lukin, \emph{Nature Physics}, 2007, \textbf{450}, 402\relax
\mciteBstWouldAddEndPuncttrue
\mciteSetBstMidEndSepPunct{\mcitedefaultmidpunct}
{\mcitedefaultendpunct}{\mcitedefaultseppunct}\relax
\EndOfBibitem
\bibitem[Falk \emph{et~al.}(2009)Falk, Koppens, Yu, Kang, de~Leon~Snapp,
  Akimov, Jo, Lukin, and Park]{Lukin2009}
A.~L. Falk, F.~H.~L. Koppens, C.~L. Yu, K.~Kang, N.~de~Leon~Snapp, A.~V.
  Akimov, M.-H. Jo, M.~D. Lukin and H.~Park, \emph{Nature Physics}, 2009,
  \textbf{10}, 475\relax
\mciteBstWouldAddEndPuncttrue
\mciteSetBstMidEndSepPunct{\mcitedefaultmidpunct}
{\mcitedefaultendpunct}{\mcitedefaultseppunct}\relax
\EndOfBibitem
\bibitem[de~Leon \emph{et~al.}(2012)de~Leon, Lukin, and Park]{Lukin2012a}
N.~P. de~Leon, M.~D. Lukin and H.~Park, \emph{IEEE Journal of Selected Topics
  in Quantum Electronics}, 2012, \textbf{18}, 1781\relax
\mciteBstWouldAddEndPuncttrue
\mciteSetBstMidEndSepPunct{\mcitedefaultmidpunct}
{\mcitedefaultendpunct}{\mcitedefaultseppunct}\relax
\EndOfBibitem
\bibitem[Shen and Fan(2007)]{Fan2007a}
J.-T. Shen and S.~Fan, \emph{Physical Review A}, 2007, \textbf{76},
  062709\relax
\mciteBstWouldAddEndPuncttrue
\mciteSetBstMidEndSepPunct{\mcitedefaultmidpunct}
{\mcitedefaultendpunct}{\mcitedefaultseppunct}\relax
\EndOfBibitem
\bibitem[Chen and Fan(2007)]{Fan2007b}
J.-T. Chen and S.~Fan, \emph{Physical Review Letters}, 2007, \textbf{98},
  153003\relax
\mciteBstWouldAddEndPuncttrue
\mciteSetBstMidEndSepPunct{\mcitedefaultmidpunct}
{\mcitedefaultendpunct}{\mcitedefaultseppunct}\relax
\EndOfBibitem
\bibitem[Shi \emph{et~al.}(2011)Shi, Fan, and Sun]{Fan2011a}
T.~Shi, S.~Fan and C.~P. Sun, \emph{Physical Review A}, 2011, \textbf{84},
  063803\relax
\mciteBstWouldAddEndPuncttrue
\mciteSetBstMidEndSepPunct{\mcitedefaultmidpunct}
{\mcitedefaultendpunct}{\mcitedefaultseppunct}\relax
\EndOfBibitem
\bibitem[Rephaeli \emph{et~al.}(2011)Rephaeli, \c{S}\"ukr\"u Kocaba\c{s}, and
  Fan]{Fan2011b}
E.~Rephaeli, \c{S}\"ukr\"u Kocaba\c{s} and S.~Fan, \emph{Physical Review A},
  2011, \textbf{84}, 063832\relax
\mciteBstWouldAddEndPuncttrue
\mciteSetBstMidEndSepPunct{\mcitedefaultmidpunct}
{\mcitedefaultendpunct}{\mcitedefaultseppunct}\relax
\EndOfBibitem
\bibitem[Xu \emph{et~al.}(2013)Xu, Rephaeli, and Fan]{Fan2013}
S.~Xu, E.~Rephaeli and S.~Fan, \emph{Physical Review Letters}, 2013,
  \textbf{111}, 223602\relax
\mciteBstWouldAddEndPuncttrue
\mciteSetBstMidEndSepPunct{\mcitedefaultmidpunct}
{\mcitedefaultendpunct}{\mcitedefaultseppunct}\relax
\EndOfBibitem
\bibitem[Shi and Sun(2009)]{Sun2009a}
T.~Shi and C.~P. Sun, \emph{Physical Review B}, 2009, \textbf{79}, 205111\relax
\mciteBstWouldAddEndPuncttrue
\mciteSetBstMidEndSepPunct{\mcitedefaultmidpunct}
{\mcitedefaultendpunct}{\mcitedefaultseppunct}\relax
\EndOfBibitem
\bibitem[Yudson and Reineker(2008)]{Reineker2008}
V.~I. Yudson and P.~Reineker, \emph{Physical Review A}, 2008, \textbf{78},
  052713\relax
\mciteBstWouldAddEndPuncttrue
\mciteSetBstMidEndSepPunct{\mcitedefaultmidpunct}
{\mcitedefaultendpunct}{\mcitedefaultseppunct}\relax
\EndOfBibitem
\bibitem[Zheng \emph{et~al.}(2010)Zheng, Gauthier, and Baranger]{Baranger2010}
H.~Zheng, D.~J. Gauthier and H.~U. Baranger, \emph{Physical Review A}, 2010,
  \textbf{82}, 063816\relax
\mciteBstWouldAddEndPuncttrue
\mciteSetBstMidEndSepPunct{\mcitedefaultmidpunct}
{\mcitedefaultendpunct}{\mcitedefaultseppunct}\relax
\EndOfBibitem
\bibitem[Zheng \emph{et~al.}(2011)Zheng, Gauthier, and Baranger]{Baranger2011}
H.~Zheng, D.~J. Gauthier and H.~U. Baranger, \emph{Physical Review Letters},
  2011, \textbf{107}, 223601\relax
\mciteBstWouldAddEndPuncttrue
\mciteSetBstMidEndSepPunct{\mcitedefaultmidpunct}
{\mcitedefaultendpunct}{\mcitedefaultseppunct}\relax
\EndOfBibitem
\bibitem[Zheng \emph{et~al.}(2012)Zheng, Gauthier, and Baranger]{Baranger2012}
H.~Zheng, D.~J. Gauthier and H.~U. Baranger, \emph{Physical Review A}, 2012,
  \textbf{85}, 043832\relax
\mciteBstWouldAddEndPuncttrue
\mciteSetBstMidEndSepPunct{\mcitedefaultmidpunct}
{\mcitedefaultendpunct}{\mcitedefaultseppunct}\relax
\EndOfBibitem
\bibitem[Zheng and Baranger(2013)]{Baranger2013}
H.~Zheng and H.~U. Baranger, \emph{Physical Review Letters}, 2013,
  \textbf{110}, 113601\relax
\mciteBstWouldAddEndPuncttrue
\mciteSetBstMidEndSepPunct{\mcitedefaultmidpunct}
{\mcitedefaultendpunct}{\mcitedefaultseppunct}\relax
\EndOfBibitem
\bibitem[Fang \emph{et~al.}(2014)Fang, Zheng, and Baranger]{Baranger2014}
Y.-L.~L. Fang, H.~Zheng and H.~U. Baranger, \emph{EPJ Quantum Technology},
  2014, \textbf{1}, 3\relax
\mciteBstWouldAddEndPuncttrue
\mciteSetBstMidEndSepPunct{\mcitedefaultmidpunct}
{\mcitedefaultendpunct}{\mcitedefaultseppunct}\relax
\EndOfBibitem
\bibitem[Roy(2011)]{Roy2011b}
D.~Roy, \emph{Physical Review A}, 2011, \textbf{83}, 043823\relax
\mciteBstWouldAddEndPuncttrue
\mciteSetBstMidEndSepPunct{\mcitedefaultmidpunct}
{\mcitedefaultendpunct}{\mcitedefaultseppunct}\relax
\EndOfBibitem
\bibitem[Roy(2013)]{Roy2013}
D.~Roy, \emph{Physical Review A}, 2013, \textbf{87}, 063819\relax
\mciteBstWouldAddEndPuncttrue
\mciteSetBstMidEndSepPunct{\mcitedefaultmidpunct}
{\mcitedefaultendpunct}{\mcitedefaultseppunct}\relax
\EndOfBibitem
\bibitem[Liao and Law(2010)]{Liao2010}
J.-Q. Liao and C.~K. Law, \emph{Physical Review A}, 2010, \textbf{82},
  053836\relax
\mciteBstWouldAddEndPuncttrue
\mciteSetBstMidEndSepPunct{\mcitedefaultmidpunct}
{\mcitedefaultendpunct}{\mcitedefaultseppunct}\relax
\EndOfBibitem
\bibitem[Peropadre \emph{et~al.}(2013)Peropadre, Lindkvist, Hoi, Wilson,
  Garc\'ia-Ripoll, Delsing, and Johansson]{Johansson2013a}
B.~Peropadre, J.~Lindkvist, I.~C. Hoi, C.~M. Wilson, J.~J. Garc\'ia-Ripoll,
  P.~Delsing and G.~Johansson, \emph{New Journal of Physics}, 2013,
  \textbf{15}, 035009\relax
\mciteBstWouldAddEndPuncttrue
\mciteSetBstMidEndSepPunct{\mcitedefaultmidpunct}
{\mcitedefaultendpunct}{\mcitedefaultseppunct}\relax
\EndOfBibitem
\bibitem[Rephaeli and Fan(2012)]{Fan2012}
E.~Rephaeli and S.~Fan, \emph{Physical Review Letters}, 2012, \textbf{108},
  143602\relax
\mciteBstWouldAddEndPuncttrue
\mciteSetBstMidEndSepPunct{\mcitedefaultmidpunct}
{\mcitedefaultendpunct}{\mcitedefaultseppunct}\relax
\EndOfBibitem
\bibitem[Chen \emph{et~al.}(2011)Chen, Lambert, Chou, Chen, and Nori]{Nori2011}
G.-Y. Chen, N.~Lambert, C.-H. Chou, Y.-N. Chen and F.~Nori, \emph{Physical
  Review B}, 2011, \textbf{84}, 045310\relax
\mciteBstWouldAddEndPuncttrue
\mciteSetBstMidEndSepPunct{\mcitedefaultmidpunct}
{\mcitedefaultendpunct}{\mcitedefaultseppunct}\relax
\EndOfBibitem
\bibitem[Gonz\'alez-Tudela \emph{et~al.}(2011)Gonz\'alez-Tudela, Cano, Moreno,
  Mart\'in-Moreno, Tejedor, and Garc\'ia-Vidal]{Tudela2011a}
A.~Gonz\'alez-Tudela, D.~M. Cano, E.~Moreno, L.~Mart\'in-Moreno, C.~Tejedor and
  F.~J. Garc\'ia-Vidal, \emph{Physical Review Letters}, 2011, \textbf{106},
  020501\relax
\mciteBstWouldAddEndPuncttrue
\mciteSetBstMidEndSepPunct{\mcitedefaultmidpunct}
{\mcitedefaultendpunct}{\mcitedefaultseppunct}\relax
\EndOfBibitem
\bibitem[Zueco \emph{et~al.}(2012)Zueco, Mazo, Solano, and
  Garc\'{\i}a-Ripoll]{Zueco2012}
D.~Zueco, J.~J. Mazo, E.~Solano and J.~J. Garc\'{\i}a-Ripoll, \emph{Physical
  Review B}, 2012, \textbf{86}, 024503\relax
\mciteBstWouldAddEndPuncttrue
\mciteSetBstMidEndSepPunct{\mcitedefaultmidpunct}
{\mcitedefaultendpunct}{\mcitedefaultseppunct}\relax
\EndOfBibitem
\bibitem[Lalumi\`{e}re \emph{et~al.}(2013)Lalumi\`{e}re, Sanders, van Loo,
  Fedorov, Wallraff, and Blais]{Wallraff2013a}
K.~Lalumi\`{e}re, B.~C. Sanders, A.~F. van Loo, A.~Fedorov, A.~Wallraff and
  A.~Blais, \emph{Physical Review A}, 2013, \textbf{88}, 043806\relax
\mciteBstWouldAddEndPuncttrue
\mciteSetBstMidEndSepPunct{\mcitedefaultmidpunct}
{\mcitedefaultendpunct}{\mcitedefaultseppunct}\relax
\EndOfBibitem
\bibitem[Gonz\'alez-Ballestero \emph{et~al.}(2014)Gonz\'alez-Ballestero,
  Moreno, and Garc\'ia-Vidal]{Ballestero2014}
C.~Gonz\'alez-Ballestero, E.~Moreno and F.~J. Garc\'ia-Vidal, \emph{Physical
  Review A}, 2014, \textbf{89}, 042328\relax
\mciteBstWouldAddEndPuncttrue
\mciteSetBstMidEndSepPunct{\mcitedefaultmidpunct}
{\mcitedefaultendpunct}{\mcitedefaultseppunct}\relax
\EndOfBibitem
\bibitem[Cohen-Tannoudji \emph{et~al.}(1992)Cohen-Tannoudji, Dupont-Roc, and
  Grynberg]{Cohen-Tannoudji1992}
C.~Cohen-Tannoudji, J.~Dupont-Roc and G.~Grynberg, \emph{{Atom-Photon
  Interactions: Basic Processes and Applications}}, Wiley-Interscience, 1992,
  p. 680\relax
\mciteBstWouldAddEndPuncttrue
\mciteSetBstMidEndSepPunct{\mcitedefaultmidpunct}
{\mcitedefaultendpunct}{\mcitedefaultseppunct}\relax
\EndOfBibitem
\bibitem[Niemczyk \emph{et~al.}(2010)Niemczyk, Deppe, Huebl, Menzel, Hocke,
  Schwarz, Garc\'{\i}a-Ripoll, Zueco, H\"{u}mmer, Solano, Marx, and
  Gross]{Niemczyk2010}
T.~Niemczyk, F.~Deppe, H.~Huebl, E.~Menzel, F.~Hocke, M.~Schwarz, J.~J.
  Garc\'{\i}a-Ripoll, D.~Zueco, T.~H\"{u}mmer, E.~Solano, A.~Marx and R.~Gross,
  \emph{Nature Physics}, 2010, \textbf{6}, 772\relax
\mciteBstWouldAddEndPuncttrue
\mciteSetBstMidEndSepPunct{\mcitedefaultmidpunct}
{\mcitedefaultendpunct}{\mcitedefaultseppunct}\relax
\EndOfBibitem
\bibitem[Forn-D\'{i}az \emph{et~al.}(2010)Forn-D\'{i}az, Lisenfeld, Marcos,
  Garc\'{i}a-Ripoll, Solano, Harmans, and Mooij]{FornDiaz2010}
P.~Forn-D\'{i}az, J.~Lisenfeld, D.~Marcos, J.~J. Garc\'{i}a-Ripoll, E.~Solano,
  C.~J. P.~M. Harmans and J.~E. Mooij, \emph{Physical Review Letters}, 2010,
  \textbf{105}, 237001\relax
\mciteBstWouldAddEndPuncttrue
\mciteSetBstMidEndSepPunct{\mcitedefaultmidpunct}
{\mcitedefaultendpunct}{\mcitedefaultseppunct}\relax
\EndOfBibitem
\bibitem[Schwartz \emph{et~al.}(2011)Schwartz, Hutchinson, Genet, and
  Ebbesen]{Schwartz2011}
T.~Schwartz, J.~A. Hutchinson, C.~Genet and T.~W. Ebbesen, \emph{Physical
  Review Letters}, 2011, \textbf{106}, 196405\relax
\mciteBstWouldAddEndPuncttrue
\mciteSetBstMidEndSepPunct{\mcitedefaultmidpunct}
{\mcitedefaultendpunct}{\mcitedefaultseppunct}\relax
\EndOfBibitem
\bibitem[G\"{u}nter \emph{et~al.}(2009)G\"{u}nter, Anappara, Hees, Sell,
  Biasol, Sorba, Liberato, Ciuti, Tredicucci, Leitenstorfer, and R]{Gunter2009}
G.~G\"{u}nter, A.~A. Anappara, J.~Hees, A.~Sell, G.~Biasol, L.~Sorba, S.~D.
  Liberato, C.~Ciuti, A.~Tredicucci, A.~Leitenstorfer and H.~R, \emph{Nature},
  2009, \textbf{458}, 07838\relax
\mciteBstWouldAddEndPuncttrue
\mciteSetBstMidEndSepPunct{\mcitedefaultmidpunct}
{\mcitedefaultendpunct}{\mcitedefaultseppunct}\relax
\EndOfBibitem
\bibitem[De~Liberato and Ciuti(2008)]{DeLiberato2008}
S.~De~Liberato and C.~Ciuti, \emph{Physical Review B}, 2008, \textbf{77},
  155321\relax
\mciteBstWouldAddEndPuncttrue
\mciteSetBstMidEndSepPunct{\mcitedefaultmidpunct}
{\mcitedefaultendpunct}{\mcitedefaultseppunct}\relax
\EndOfBibitem
\bibitem[Hur(2012)]{Lehur2012}
K.~L. Hur, \emph{Physical Review B}, 2012, \textbf{85}, 140506(R)\relax
\mciteBstWouldAddEndPuncttrue
\mciteSetBstMidEndSepPunct{\mcitedefaultmidpunct}
{\mcitedefaultendpunct}{\mcitedefaultseppunct}\relax
\EndOfBibitem
\bibitem[Romero \emph{et~al.}(2012)Romero, Ballester, Wang, Scarani, and
  Solano]{Romero2012}
G.~Romero, D.~Ballester, Y.~M. Wang, V.~Scarani and E.~Solano, \emph{Phys. Rev.
  Lett.}, 2012, \textbf{108}, 120501\relax
\mciteBstWouldAddEndPuncttrue
\mciteSetBstMidEndSepPunct{\mcitedefaultmidpunct}
{\mcitedefaultendpunct}{\mcitedefaultseppunct}\relax
\EndOfBibitem
\bibitem[Peropadre \emph{et~al.}(2013)Peropadre, Zueco, Porras, and
  Garc\'{\i}a-Ripoll]{Peropadre2013}
B.~Peropadre, D.~Zueco, D.~Porras and J.~J. Garc\'{\i}a-Ripoll, \emph{Physical
  Review Letters}, 2013, \textbf{111}, 243602\relax
\mciteBstWouldAddEndPuncttrue
\mciteSetBstMidEndSepPunct{\mcitedefaultmidpunct}
{\mcitedefaultendpunct}{\mcitedefaultseppunct}\relax
\EndOfBibitem
\bibitem[Naether \emph{et~al.}(2014)Naether, Garc\'ia-Ripoll, Mazo, and
  Zueco]{Naether2014}
U.~Naether, J.~J. Garc\'ia-Ripoll, J.~J. Mazo and D.~Zueco, \emph{Phys. Rev.
  Lett.}, 2014, \textbf{112}, 074101\relax
\mciteBstWouldAddEndPuncttrue
\mciteSetBstMidEndSepPunct{\mcitedefaultmidpunct}
{\mcitedefaultendpunct}{\mcitedefaultseppunct}\relax
\EndOfBibitem
\bibitem[S\'anchez-Burillo \emph{et~al.}(2014)S\'anchez-Burillo, Zueco,
  Garc\'ia-Ripoll, and Mart\'in-Moreno]{Burillo2014}
E.~S\'anchez-Burillo, D.~Zueco, J.~J. Garc\'ia-Ripoll and L.~Mart\'in-Moreno,
  \emph{arXiv:1406.5779v1 [quant-ph]}, 2014\relax
\mciteBstWouldAddEndPuncttrue
\mciteSetBstMidEndSepPunct{\mcitedefaultmidpunct}
{\mcitedefaultendpunct}{\mcitedefaultseppunct}\relax
\EndOfBibitem
\bibitem[Wang \emph{et~al.}(2012)Wang, Li, Zhou, Sun, and Zhang]{Sun2012}
Z.~H. Wang, Y.~Li, D.~L. Zhou, C.~P. Sun and P.~Zhang, \emph{Physical Review
  A}, 2012, \textbf{86}, 023824\relax
\mciteBstWouldAddEndPuncttrue
\mciteSetBstMidEndSepPunct{\mcitedefaultmidpunct}
{\mcitedefaultendpunct}{\mcitedefaultseppunct}\relax
\EndOfBibitem
\bibitem[Dicke(1954)]{Dicke1954}
R.~H. Dicke, \emph{Physical Review}, 1954, \textbf{93}, 99\relax
\mciteBstWouldAddEndPuncttrue
\mciteSetBstMidEndSepPunct{\mcitedefaultmidpunct}
{\mcitedefaultendpunct}{\mcitedefaultseppunct}\relax
\EndOfBibitem
\bibitem[Witthaut and S{\o}rensen(2010)]{Sorensen2010}
D.~Witthaut and A.~S. S{\o}rensen, \emph{New Journal of Physics}, 2010,
  \textbf{12}, 043052\relax
\mciteBstWouldAddEndPuncttrue
\mciteSetBstMidEndSepPunct{\mcitedefaultmidpunct}
{\mcitedefaultendpunct}{\mcitedefaultseppunct}\relax
\EndOfBibitem
\bibitem[Roy(2011)]{Roy2011a}
D.~Roy, \emph{Physical Review Letters}, 2011, \textbf{106}, 053601\relax
\mciteBstWouldAddEndPuncttrue
\mciteSetBstMidEndSepPunct{\mcitedefaultmidpunct}
{\mcitedefaultendpunct}{\mcitedefaultseppunct}\relax
\EndOfBibitem
\bibitem[Cohen-Tannoudji \emph{et~al.}(1997)Cohen-Tannoudji, Dupont-Roc, and
  Grynberg]{Cohen-Tannoudji1997}
C.~Cohen-Tannoudji, J.~Dupont-Roc and G.~Grynberg, \emph{{Photons \& Atoms:
  Introduction to Quanutm Electrodynamics}}, Wiley Vch, 1997, p. 486\relax
\mciteBstWouldAddEndPuncttrue
\mciteSetBstMidEndSepPunct{\mcitedefaultmidpunct}
{\mcitedefaultendpunct}{\mcitedefaultseppunct}\relax
\EndOfBibitem
\bibitem[Vidal(2003)]{Vidal2003}
G.~Vidal, \emph{Physical Review Letters}, 2003, \textbf{91}, 147902\relax
\mciteBstWouldAddEndPuncttrue
\mciteSetBstMidEndSepPunct{\mcitedefaultmidpunct}
{\mcitedefaultendpunct}{\mcitedefaultseppunct}\relax
\EndOfBibitem
\bibitem[Vidal(2004)]{Vidal2004}
G.~Vidal, \emph{Physical Review Letters}, 2004, \textbf{93}, 040502\relax
\mciteBstWouldAddEndPuncttrue
\mciteSetBstMidEndSepPunct{\mcitedefaultmidpunct}
{\mcitedefaultendpunct}{\mcitedefaultseppunct}\relax
\EndOfBibitem
\bibitem[Verstraete \emph{et~al.}(2008)Verstraete, Murg, and
  Cirac]{Verstraete2008}
F.~Verstraete, V.~Murg and J.~I. Cirac, \emph{Advances in Physics}, 2008,
  \textbf{57}, 143\relax
\mciteBstWouldAddEndPuncttrue
\mciteSetBstMidEndSepPunct{\mcitedefaultmidpunct}
{\mcitedefaultendpunct}{\mcitedefaultseppunct}\relax
\EndOfBibitem
\bibitem[Eisert \emph{et~al.}(2010)Eisert, Cramer, and Plenio]{Eisert2010}
J.~Eisert, M.~Cramer and M.~B. Plenio, \emph{Rev. Mod. Phys.}, 2010,
  \textbf{82}, 277--306\relax
\mciteBstWouldAddEndPuncttrue
\mciteSetBstMidEndSepPunct{\mcitedefaultmidpunct}
{\mcitedefaultendpunct}{\mcitedefaultseppunct}\relax
\EndOfBibitem
\bibitem[Suzuki(1991)]{Suzuki1991}
M.~Suzuki, \emph{Journal of Mathematical Physics}, 1991, \textbf{32}, 400\relax
\mciteBstWouldAddEndPuncttrue
\mciteSetBstMidEndSepPunct{\mcitedefaultmidpunct}
{\mcitedefaultendpunct}{\mcitedefaultseppunct}\relax
\EndOfBibitem
\bibitem[Garc\'ia-Ripoll(2006)]{JJRipoll2006}
J.~J. Garc\'ia-Ripoll, \emph{New Journal of Physics}, 2006, \textbf{8},
  305\relax
\mciteBstWouldAddEndPuncttrue
\mciteSetBstMidEndSepPunct{\mcitedefaultmidpunct}
{\mcitedefaultendpunct}{\mcitedefaultseppunct}\relax
\EndOfBibitem
\bibitem[H\"{u}mmer \emph{et~al.}(2012)H\"{u}mmer, Reuther, H\"{a}nggi, and
  Zueco]{Hummer2012}
T.~H\"{u}mmer, G.~Reuther, P.~H\"{a}nggi and D.~Zueco, \emph{Physical Review
  A}, 2012, \textbf{85}, 052320\relax
\mciteBstWouldAddEndPuncttrue
\mciteSetBstMidEndSepPunct{\mcitedefaultmidpunct}
{\mcitedefaultendpunct}{\mcitedefaultseppunct}\relax
\EndOfBibitem
\bibitem[Shen and Fan(2005)]{Fan2005a}
J.-T. Shen and S.~Fan, \emph{Optics Letters}, 2005, \textbf{30}, 2001\relax
\mciteBstWouldAddEndPuncttrue
\mciteSetBstMidEndSepPunct{\mcitedefaultmidpunct}
{\mcitedefaultendpunct}{\mcitedefaultseppunct}\relax
\EndOfBibitem
\bibitem[Shen and Fan(2005)]{Fan2005b}
J.-T. Shen and S.~Fan, \emph{Physical Review Letters}, 2005, \textbf{95},
  213001\relax
\mciteBstWouldAddEndPuncttrue
\mciteSetBstMidEndSepPunct{\mcitedefaultmidpunct}
{\mcitedefaultendpunct}{\mcitedefaultseppunct}\relax
\EndOfBibitem
\bibitem[Zhou \emph{et~al.}(2008)Zhou, Gong, xi~Liu, Sun, and Nori]{Nori2008a}
L.~Zhou, Z.~R. Gong, Y.~xi~Liu, C.~P. Sun and F.~Nori, \emph{Physical Review
  Letters}, 2008, \textbf{101}, 100501\relax
\mciteBstWouldAddEndPuncttrue
\mciteSetBstMidEndSepPunct{\mcitedefaultmidpunct}
{\mcitedefaultendpunct}{\mcitedefaultseppunct}\relax
\EndOfBibitem
\bibitem[Zhou \emph{et~al.}(2009)Zhou, Yang, xi~Liu, Sun, and Nori]{Nori2009}
L.~Zhou, S.~Yang, Y.~xi~Liu, C.~P. Sun and F.~Nori, \emph{Physical Review A},
  2009, \textbf{80}, 062109\relax
\mciteBstWouldAddEndPuncttrue
\mciteSetBstMidEndSepPunct{\mcitedefaultmidpunct}
{\mcitedefaultendpunct}{\mcitedefaultseppunct}\relax
\EndOfBibitem
\bibitem[Liao \emph{et~al.}(2010)Liao, Gong, Zhou, xi~Liu, Sun, and
  Nori]{Nori2010}
J.-Q. Liao, Z.~R. Gong, L.~Zhou, Y.~xi~Liu, C.~P. Sun and F.~Nori,
  \emph{Physical Review A}, 2010, \textbf{81}, 042304\relax
\mciteBstWouldAddEndPuncttrue
\mciteSetBstMidEndSepPunct{\mcitedefaultmidpunct}
{\mcitedefaultendpunct}{\mcitedefaultseppunct}\relax
\EndOfBibitem
\bibitem[Zhou \emph{et~al.}(2013)Zhou, Yang, Li, and Sun]{Zhou2013}
L.~Zhou, L.-P. Yang, Y.~Li and C.~P. Sun, \emph{Physical Review Letters}, 2013,
  \textbf{111}, 103604\relax
\mciteBstWouldAddEndPuncttrue
\mciteSetBstMidEndSepPunct{\mcitedefaultmidpunct}
{\mcitedefaultendpunct}{\mcitedefaultseppunct}\relax
\EndOfBibitem
\bibitem[Lu \emph{et~al.}(2014)Lu, Zhou, Kuang, and Nori]{Lu2014}
J.~Lu, L.~Zhou, L.-M. Kuang and F.~Nori, \emph{Physical Review A}, 2014,
  \textbf{89}, 013805\relax
\mciteBstWouldAddEndPuncttrue
\mciteSetBstMidEndSepPunct{\mcitedefaultmidpunct}
{\mcitedefaultendpunct}{\mcitedefaultseppunct}\relax
\EndOfBibitem
\bibitem[Lehmberg(1970)]{Lehmberg1970}
R.~H. Lehmberg, \emph{Physical Review A}, 1970, \textbf{2}, 883\relax
\mciteBstWouldAddEndPuncttrue
\mciteSetBstMidEndSepPunct{\mcitedefaultmidpunct}
{\mcitedefaultendpunct}{\mcitedefaultseppunct}\relax
\EndOfBibitem
\bibitem[Zhou \emph{et~al.}(2008)Zhou, Dong, xi~Liu, Sun, and Nori]{Nori2008b}
L.~Zhou, H.~Dong, Y.~xi~Liu, C.~P. Sun and F.~Nori, \emph{Physical Review A},
  2008, \textbf{78}, 063827\relax
\mciteBstWouldAddEndPuncttrue
\mciteSetBstMidEndSepPunct{\mcitedefaultmidpunct}
{\mcitedefaultendpunct}{\mcitedefaultseppunct}\relax
\EndOfBibitem
\bibitem[Longo \emph{et~al.}(2010)Longo, Schmitteckert, and Busch]{Longo2010}
P.~Longo, P.~Schmitteckert and K.~Busch, \emph{Physical Review Letters}, 2010,
  \textbf{104}, 023602\relax
\mciteBstWouldAddEndPuncttrue
\mciteSetBstMidEndSepPunct{\mcitedefaultmidpunct}
{\mcitedefaultendpunct}{\mcitedefaultseppunct}\relax
\EndOfBibitem
\bibitem[Longo \emph{et~al.}(2011)Longo, Schmitteckert, and Busch]{Longo2011}
P.~Longo, P.~Schmitteckert and K.~Busch, \emph{Physical Review A}, 2011,
  \textbf{83}, 063828\relax
\mciteBstWouldAddEndPuncttrue
\mciteSetBstMidEndSepPunct{\mcitedefaultmidpunct}
{\mcitedefaultendpunct}{\mcitedefaultseppunct}\relax
\EndOfBibitem
\bibitem[Shi and Sun(2009)]{Sun2009b}
T.~Shi and C.~P. Sun, \emph{arXiv:0907.2776 [quant-ph]}, 2009\relax
\mciteBstWouldAddEndPuncttrue
\mciteSetBstMidEndSepPunct{\mcitedefaultmidpunct}
{\mcitedefaultendpunct}{\mcitedefaultseppunct}\relax
\EndOfBibitem
\end{mcitethebibliography}
\bibliographystyle{rsc}

}

\end{document}